\documentclass[a4paper,11 pt,english]{article}
\usepackage[T1]{fontenc}
\usepackage{times}

\usepackage{amsmath,amssymb,amsfonts}
\usepackage{graphicx} 
\usepackage{caption, subcaption}
\usepackage{abstract}

\usepackage{physics}
\usepackage{upquote}
\usepackage[margin=2.5cm]{geometry}
\usepackage[onehalfspacing]{setspace}
\usepackage{cite}
\usepackage{hyperref}
\usepackage[affil-it]{authblk}
\usepackage[dvipsnames]{xcolor}
\usepackage{sectsty}

\title{
	Origin of selective enhancement of sharp defect emission lines in monolayer WSe$_2$ on rough metal substrate}
\author{Raghav Chaudhary, Varun Raghunathan, Kausik Majumdar$^*$\\
	Department of Electrical Communication Engineering, Indian Institute of Science, Bangalore 560012, India \\
	$^*$Corresponding author, E-mail: kausikm@iisc.ac.in}

\date{}

\begin{document}
	\maketitle
	\begin{abstract}
	The defect states in atomically thin layers of transition metal dichalcogenides are promising candidates for single photon emission. However, the brightness of such quantum emission is often weak, and is accompanied with undesirable effects like spectral diffusion and strong background emission. By placing a monolayer WSe$_2$ directly on a rough gold substrate, here we show a selective enhancement of sharp defect-bound exciton peaks, coupled with a suppressed spectral diffusion and strong quenching of background luminescence. By combining the experimental data with detailed electromagnetic simulations, we reveal that such selective luminescence enhancement originates from a combination of the Purcell effect and a wavelength dependent increment of the excitation electric field at the tips of tall rough features, coupled with a localized strain induced exciton funneling effect. Notably, insertion of a thin hexagonal Boron Nitride (hBN) sandwich layer between WSe$_2$ and the Au film results in a strong enhancement of the background luminescence, obscuring the sharp defect peaks. The findings demonstrate a simple strategy of using monolayer WSe$_2$ supported by thin metal film that offers a possibility of achieving quantum light sources with high purity, high brightness, and suppressed spectral diffusion.
	\end{abstract}
	\pagebreak
	\renewcommand{\thesection}{\arabic{section}}
	\renewcommand{\thesubsection}{\thesection.\arabic{subsection}}
	
	\section{INTRODUCTION}
	
	The evolving field of quantum technology aims at utilizing the intriguing principles of quantum mechanics in order to achieve faster computation, secure communication protocols, more precise measurements and simulation of quantum systems\cite{Browne}. A key element in the implementation of quantum technology is the single photon emitter (SPE).  
	SPEs today find applications in quantum computation, quantum cryptography\cite{Beveratos2002} and quantum metrology\cite{Giovannetti2011,Motes2015}. A plethora of SPE systems have been studied to date such as trapped atoms and ions, colour centres in high band gap materials like diamond and nanostructures such as carbon nanotubes and quantum dots\cite{Aharonovich2016}. Signatures of SPEs have also been observed from highly localized spots close to the edges of the flakes of atomically thin layers of transition metal dichalcogenides (TMDs)\cite{Koperski2014a,Srivastava2014,Chakraborty2015c,Tonndorf2015,He2015b,Palacios-Berraquero2016,Clark2016} and also from other 2D materials such as graphene\cite{Zhao2018} and hexagonal Boron Nitride (hBN)\cite{Tran2016,Tran2016c,Grosso2017,Bourrellier2016,Ziegler2019}. The microscopic origin of these SPEs is yet to be thoroughly understood, but experimental results\cite{Chakraborty2015c,He2015b} and theoretical calculations\cite{Zheng2019,Tran2016,Abdi2018,Tran2016c} attribute their origin to the recombination of excitons bound to quantum dot like confinement potentials which arise from unique point defects in the crystal. SPEs in 2D materials also offer the possibility of facile integration with waveguides, and thus pave the way for the implementation of on-chip quantum photonic circuits\cite{Tonndorf2017,Blauth2018}.	
	
	\par Notwithstanding their utility as quantum emitters, monolayer (1L) TMDs suffer from a number of limitations, when compared to well known quantum light emitters in high band gap materials such as diamond and even hBN. For one, the quantum emission is observed only at cryogenic temperatures. The natural occurrence of these emitters is random and restricted mostly to spots close to the edges of the flake. The brightness of these emitters is also limited to about an order of magnitude lower than hBN. Moreover, excitons bound to point defects in these atomically thin flakes are subject to interactions with traps in the underlying substrate and the surrounding defects in the material, which leads to a random jittering or "spectral diffusion" of the emission lines, up to a few meV on the energy scale \cite{Chakraborty2015c,Tonndorf2015,Srivastava2014,Koperski2014a}. The spectral diffusion effect is detrimental from the standpoint of achieving indistinguishable single photon emission, which is an indispensable requirement for applications in various quantum technological applications. Another crucial limitation is the pronounced background emission from 1L-TMDs, consisting of the luminescence from the free excitons, trions and biexcitons{Huang2016}. This background emission degrades the "purity" of quantum emission in 1L-TMDs. Various design strategies have been previously employed to overcome these limitations. For example, Branny et al.\cite{branny_deterministic_2017} and Palacios-Berraquero et al. \cite{Palacios-Berraquero2017} transferred 1L-WSe$_2$ on top of lithographically defined nanopillars. The nanopillars serve to locally strain the WSe$_2$ flake, which not only creates point defects but also leads to a local strain-induced dip in the band gap, thus causing a trapping or funneling of free excitons from the vicinity into the defects introduced by the nanopillars. Recent works by various groups\cite{Johnson2017,Tripathi2018a,Cai2018,Luo,Luo2019,Iff:18} have also achieved a deterministic and enhanced quantum emission in 1L-WSe$_2$ by coupling the defect peak emission to the localized surface plasmon resonances (LSPRs) of various plasmonic nanostructures. 
	\par In this work, we report a selective enhancement of multiple sharp defect PL peaks in 1L-WSe$_2$ coupled with a quenching of the free exciton peak, when placed directly on top of a rough gold (Au) film, with no dielectric spacer in between. Such a selective enhancement employing this simple system is an attractive feature for this system to be used as a viable option for quantum light emission. Using a combination of experimental data and detailed electromagnetic simulations, we attribute the free exciton quenching to fast energy and charge transfer from 1L-WSe$_2$ to Au. On the other hand, the roughness of the metal film surface, introduced during the fabrication process, plays a crucial role in the enhancement of the defect luminescence. The electromagnetic simulations show that the selective enhancement is achieved through a combination of the Purcell effect and a wavelength dependent enhancement of the excitation electric field at the tips of the rough features on the metal substrate. Apart from the electromagnetic effects, the localized strain introduced into the flake when it conforms to the roughness features can also lead to a preferential population of the point defects associated with these features by the exciton funnelling effect. Importantly, the Gaussian component of the full-width at half-maximum (FWHM) of the enhanced defect peaks reduces significantly on the metal film - a signature of suppressed spectral diffusion, which we attribute to the screening of charge fluctuations in the SiO$_2$ substrate by the metal film. On the other hand, sandwiching a thin hBN layer between the WSe$_2$ and the Au film enhances the background PL, and obscures the sharp defect peaks. The above findings corroborate the potential role of directly transferring atomically thin flakes of WSe$_2$ on patterned metallic substrates in the formation and enhancement of quantum emitters in atomically thin materials, and also a simple strategy that can improve their emission properties.
	
	\section{EXPERIMENT}	
	50 nm thick and 5 $\mu$m wide Au lines are patterned on $285$ nm SiO$_2$ coated Si substrate. To fabricate the Au lines, close to 300 nm of Polymethyl methacrylate (PMMA) is spin-coated on the substrate and the line patterns are defined by electron beam lithography. Post development, Ni/Au (10 nm/50 nm) is deposited at room temperature in an Argon atmosphere at a working pressure of 6 mTorr via DC magnetron sputtering. After the metal liftoff in acetone, 1L-WSe$_2$ flakes are mechanically exfoliated from bulk crystals (2D Semiconductors) onto a Polydimethylsiloxane (PDMS) stamp and subsequently dry-transferred \cite{Castellanos-Gomez2014} on the pre-patterned substrate such that some portion of the flake lies on top of the Au line, and the remaining portion is on SiO$_2$, which provides for a control. After transferring, the substrate was heated at 80 degree Celsius for 2 minutes on a hot plate under ambient atmospheric conditions. The heating plays an important role in overcoming the "Van der Waals gap" intrinsic to vertical heterostructures of van Der Waal's materials, thus ensuring a better contact between the Au film and the 1L-WSe$_2$ flake. The optical micrograph of the flake is shown in Figure 1a. The edge of the monolayer flake has been highlighted with the white dashed lines The flake thickness is confirmed through Raman spectroscopy at an excitation wavelength of 532 nm. As shown in Figures 1b, the degenerate E$^1_{2g}$ and A$_{1g}$ modes are observed at a stokes shift of 251 cm$^{-1}$ on SiO$_2$/Si substrate. Notably, the B$_{2g}$ peak at 309 cm$^{-1}$ is not observed on either SiO$_2$ or Au substrate. This unambiguously establishes the monolayer nature of the flake\cite{Zhao2013}. 
	
	\par Having confirmed the monolayer nature of the WSe$_2$ flake on the SiO$_2$ substrate and metal film, photoluminescence is acquired from different locations of the 1L-flake on both the substrates. The sample is placed in a liquid He closed cycle cryostat with an optical window above the sample stage, and the temperature is maintained at 3.7 K. The flake is excited at 532 nm and 633 nm wavelengths using a confocal microscope. A 50$\times$ magnification, 0.5 numerical aperture (NA) objective is used to focus the excitation laser light into a diffraction limited spot. The emission from the flake is collected by the same objective and passed through a long-pass filter, which rejects the scattered excitation laser and directs the luminescence to a spectrometer.
	
	\section{RESULTS AND DISCUSSION}
	\subsection{Selective enhancement of narrow defect peaks}
	\par Figure 1c shows a representative PL spectrum obtained by exciting the flake on the two substrates (namely, SiO$_2$ and Au) with a continuous wave 532 nm laser at 8 $\mu$W power (more spectra at 532 nm and 633 nm excitations taken from different locations on the same flake are shown in supplementary section S1). All PL spectra have been acquired over a 30 seconds integration time. The respective locations on the flake from where each spectrum is collected are represented by the dots in the optical image in the inset. The photoluminescence peaks labelled as X$^0$ correspond to the free exciton peak, as reported in previous low temperature PL studies on 1L-WSe$_2$\cite{Huang2016,Clark2016,He2015b}. Apart from a quenching of the free exciton peak, we repeatedly observe a small blue-shift of the free exciton peak on the Au film. This shift arises from an interplay between the dielectric screening and the bandgap renormalization effect on the Au substrate\cite{Ugeda2014a,Chernikov2014,Gupta2017a}. We also consistently observe a large number of enhanced and sharp peaks, mostly in the range of 1.6 to 1.7 eV, when 1L-WSe$_2$ is on Au. 
	To determine the origin of these sharp enhanced peaks, we examine the relation between the intensity of these features as a function of laser excitation power. The excitation power dependent peak intensities are often described by a power law dependence with a single exponent, the value of which characterizes the origin of the peaks\cite{Schmidt1992}. Figure 1d shows a representative plot for the peaks labelled as D1 and D2 in figure 1c, as well as the free exciton peaks X$^0$. A linear fitting (equivalent to a power law fit in the linear scale) for peaks D1 and D2 for the range of excitation powers considered in this experiment returns a slope (and therefore, a power law exponent) $\alpha$ of 0.69 and 0.55 for WSe$_2$ on SiO$_2$/Si and Au substrates respectively. Thus, the origin of these sharp features can be attributed to defect related transitions\cite{Schmidt1992,Huang2016}. The slope corresponding to the X$^0$ peak is extracted at 0.98 and 1.16 on SiO$_2$/Si and metal substrates respectively, which confirms their origin from the free exciton radiative recombination\cite{Schmidt1992,Huang2016}. Previous low temperature PL studies on 1L-WSe$_2$ have consistently reported the peaks falling in the range of 1.6 to 1.7 eV as the defect peaks\cite{Clark2016}. Therefore, in this work, we attribute the enhanced and sharp features on the Au film to arise from radiative recombination from point defects. It is also shown by means of excitation power dependent PL spectra (from the same locations) in supplementary section S2 that the peaks corresponding to the trion and the biexcitons also quench when the WSe$_2$ is directly transferred on the metal film.

	The appearance of the sharp defect features is correlated with the degree of contact between the 1L-WSe$_2$ flake and the Au film, as demonstrated in supplementary section S3. This is evidenced by the observation that the sharp defect peaks are conspicuous for the 1L-WSe$_2$ flake on the Au film only after the substrate has been heated to ensure better contact between the Au film and the monolayer flake. When the flake is not heated, the sharp features are not observed due to the van der Waal's gap that exists between the flake and the substrate. 
	
	The rate of spontaneous emission of a dipole emitter is not just a material property, but is also dependent on the environment in which the dipole emitter is located. For example, in a cavity or in the vicinity of a plasmonic nanostructure, the rate of spontaneous emission can be enhanced, which is known as the Purcell effect\cite{Purcell1946}, and it arises because of an enhancement of the local density of electromagnetic states contributed by radiative modes. Of interest in the context of this work is the influence of a metal film on the total decay rate of the dipole emitter. This has been investigated both theoretically and experimentally by the pioneering works of Chance et al\cite{Chance1974,Chance2007}. Classically speaking, the dipole's decay rate is influenced by the fact that the field reflected by the metal mirror interacts with the oscillating dipole, and depending on the phase difference between the dipole oscillation and the scattered field, the dipole oscillations are either driven more strongly or damped further. Chance et al.\cite{Chance1975a} further determined that very close to the metallic surface, the dipole's near-field emission can resonantly couple to the surface plasmon modes of the metal mirror, thus facilitating a non-radiative energy transfer from the dipole to the metal. Thus, the presence of the metal mirror can influence the radiative and non-radiative decay rates of the dipole emitter, leading to a modification of the emission quantum yield.

As the 1L-WSe$_2$ is placed right on top of the Au film in this experiment, the exciton dipoles, both free and bound, will be located extremely close to the Au film surface. Based on the discussion in the previous paragraph, a significant loss of the excitons due to energy transfer from 1L-WSe$_2$ into the Au is expected. Also, the band alignment between Au and 1L-WSe$_2$ favours a charge transfer of excitons into the Au film. The free exciton quenching, as experimentally observed in Figure 1c, is in agreement with the above phenomena, but the selective enhancement of the defect peaks must be accounted by an independent effect that compensates for the reduced emission quantum yield. We note that the experimental results pertaining to the selective enhancement of the defect luminescence of 1L-WSe$_2$ are also reported in some previous works\cite{Tripathi2018a,Johnson2017}, where 1L-WSe$_2$ is transferred onto plasmonic nanostructures. In this work, the surface of the sputtered metal film, on account of a finite roughness can host such nanostructures, albeit with a statistical distribution of sizes. To verify the same, we obtain a histogram (Figure 2a) of the roughness feature heights for the portions of the Au film supporting the 1L-WSe$_2$ using Atomic Force Microscopy (AFM). The rough metal surface implies that as the WSe$_2$ film wraps over the rough features, the portion of the flake close to the tips of the features experiences a tensile strain, which can give rise to point defects in the 1L-WSe$_2$ film. Tall roughness features can also enhance the excitation laser field at their tips, due to a combination of the surface plasmon excitation and a purely geometrical factor termed as the lightning rod effect\cite{Liao1982}. The resonant excitation of the surface plasmons can enhance the radiative decay rate of the defect excitons (the Purcell effect). The lightning rod effect, on the other hand, entails that features with high aspect ratio (height$/$radius) can give high excitation field enhancement with the enhanced field concentrated at the tips, which effectively increases the rate at which the defect excitons are being generated. The overall PL intensity for an exciton dipole is defined as
	\begin{equation}
	I_{PL} = |\textbf{p}.\textbf{E}|^2 \frac{\gamma_R}{\gamma_R + \gamma_{NR}}
	\end{equation}
	where the $|\textbf{p}.\textbf{E}|^2$ term represents the square magnitude of the dot product of the excitation laser electric field and the exciton dipole moment vector, and the expression $\eta = \frac{\gamma_R}{\gamma_R + \gamma_{NR}}$, where $\gamma_R$ refers to the radiative decay rate of the exciton and $\gamma_{NR}$ refers to the non-radiative decay rate, defines the emission quantum yield ($\eta$). $\eta$ can also be expressed in terms of the electromagnetic power radiated by the dipole into the far field and the power absorbed by the underlying substrate as\cite{Bharadwaj2009}(see also supplementary information section S2.1)
	\begin{equation}
	\eta = \frac{\frac{P_{rad}}{P_0}}{\frac{P_{rad}}{P_0} + \frac{P_{loss}}{P_0} + \frac{(1-\eta_i)}{\eta_i}}
	\end{equation}
where P$_{rad}$ denotes the power radiated by the dipole into the far field, P$_0$ refers to the total power radiated by the dipole in free space, P$_{loss}$ denotes the power absorbed by the substrates and $\eta_i$ refers to the intrinsic quantum yield of 1L-WSe$_2$ (assumed to be 0.1\%).

	\par \textbf{Role of Purcell effect:} We first discuss the effect of the roughness features on $\eta$ for the defect exciton through electromagnetic simulations, using the Finite Difference Time Domain (FDTD) methodology. The simulation details for the calculation of $\eta$ are outlined in supplementary section S4.1, and the schematic is illustrated in figure 2b. Briefly, the rough protrusion is modeled as an ellipsoid with the base diameter $a = 2$ nm and height ($h$, see figure 2b) varied. The defect bound exciton is modeled as a dipole emitter, radiating at 753 nm (1.65 eV), 4 {\AA} above the tip of the ellipsoid,inclined at an angle of 45 degrees with respect to the major axis of the ellipsoid. The choice of 45 degrees stems from the fact that the experimentally obtained defect PL intensity will include all possible dipole orientations with respect to the electric field vector at the dipole's location, as is indicated in the dot product in equation (1). Thus, the total defect PL intensity is obtained by integrating equation (1) over all dipole orientations with respect to the electric field vector, i.e. from $\theta = 0$ to $\theta = \pi$. $I_{PL-net} \propto \int_{0}^{\pi} \cos[2](\theta) d\theta$. The integral, assuming an emission quantum yield independent of the dipole orientation, evaluates to $\pi/2$. Also, the PL intensty arising from all the possible defect dipole orientations can be effectively modeled as a single dipole with a fixed orientation such that $\int_{0}^{\pi}$ $\cos[2](\theta_{eff}) = \pi/2$, which implies that $\pi\cos[2](\theta_{eff}) = \pi/2$ and therefore $\theta_{eff} = 45$ degree.
	Also, it may be argued that a rough surface be modeled as an array or distribution of such ellipsoidal features, we emphasize that the exciton will be bound to point defects localized to a single ellipsoid only by virtue of the exciton funneling effect, which shall be elaborated upon. It is also verified through the surface roughness plots that the rough features are sufficiently far away so as to preclude the role of the interaction between neighbouring features in the enhancement of the electric field (also termed as "hot-spots"). Therefore, in the context of the defect bound exciton, we find it reasonable to consider a dipole close to the tip of a single ellipsoid. The radiative decay rate is quantified as the power radiated into the far field by the dipole, and the non-radiative decay rate is quantified as the absorption losses in the underlying substrates (see the discussion leading to equation (2)). The radiative and the non-radiative decay rates, normalized with respect to the same quantities for a dipole emitter (753 nm) inclined at $\theta_{eff} = 45\deg$ and 4\AA above a smooth SiO$_2$/Si substrate, define the enhancement of the radiative and non-radiative decay rates for the exciton dipole on Au with respect to the SiO$_2$/Si substrate, and are shown in figures 2c and 2d respectively, as a function of the Au ellipsoid aspect ratio $2h/a$. The results predict a significant increase in both the radiative and non-radiative decay rate. Both the radiative and non-radiative decay rates reach a maximum at an aspect ratio of 15. The drastic increase at this aspect ratio can be attributed to the resonant excitation of localized surface plasmons for the simulated structure, which results in a drastic increase in the total power dissipated (total decay rate) by the dipole. The 1L-WSe$_2$ is right on top of the Au film, and the atomically thin monolayer ensures that a significant portion of the dipole's dissipated energy is coupled to non-radiative losses in the metal film\cite{Chance1975a}, which explains the increase in the non-radiative decay rate for the dipole emitter close to the metal, compared to the same emitter on a smooth SiO$_2$ substrate. Interestingly, the simulations also predict an enhancement in the radiative decay rate, which means that the Purcell effect is also playing a role. Considering that the intrinsic quantum yield $\eta_i$ is small for the monolayer TMD, even if the non-radiative decay rate increases significantly in the presence of the metal ellipsoid, the denominator term in the expression of $\eta$ (see equation 2) does not increase as significantly as the radiative decay rate or the numerator. This implies that compared to a dipole emitter close to the SiO$_2$/Si substrate, $\eta$ for the same dipole close to the Au film can enhance, as shown in the plot of figure 2e. However, it should also be noted that the EM simulations do not account for loss of excitons due to charge transfer into the metal. The charge transfer, qualitatively speaking, can be modeled as an additional power loss term in the denominator, but it is expected not to significantly pull down the emission quantum yield. The above analysis clearly shows that the Purcell effect can play a role in the defect PL enhancement of 1L-WSe$_2$ directly on top of a Au film, despite the close proximity of the dipole to the film.
	
	\par \textbf{Role of excitation field enhancement and localized strain:} Next, the role of the excitation field enhancement is investigated by means of FDTD simulations. To emulate the laser light emanating from the microscope objective in the experiment, the excitation source in the simulation is modeled as a Gaussian beam emanating from a lens of numerical aperture (NA) 0.5. The beam is focused on the surface of the Au film. Excitations at 532 nm and 633 nm are considered. The simulation details are outlined in supplementary section S4.2. It is known that the electric field in the focal plane of a lens can also carry energy in the longitudinal direction (along the direction of propagation)\cite{Novotny2012}. This longitudinal component is important, as it will be oriented along the major axis of the ellipsoid, and will contribute to the field enhancement at the tip\cite{Liao1982}. So, we first consider the electric field profile of the Gaussian beam on the Au surface, and place the ellipsoid at the point where the longitudinal component of the electric field (along z-axis) is maximum. The electric field intensity enhancement factor $(|E|/|E_0|)^2$, where $E$ refers to the electric field amplitude 4\AA above the tip of the ellipsoid and E$_0$ denotes the excitation electric field amplitude, is plotted on the left axis as a function of the aspect ratio $2h/a$ in Figure 3a-b for 532 and 633 nm excitation. The field enhancement factor is higher for 633 nm excitation than 532 nm, due to the higher absorption of Au at 532 nm \cite{Wang2016}. Note that the region of enhanced electric field is restricted to a small region close to the tips of the ellipsoid, as shown in the representative color plot of the electric field intensity on the Au film in Figure 3c and also in supplementary section S4.2, figure S4c. Though some field enhancement occurs at the lower corners as well, it is difficult to argue if the WSe$_2$ flake will actually come in contact with that portion. If a defect in 1L-WSe$_2$ happens to occur on the tip of the feature, the enhanced excitation field experienced by the defect excitons can also compensate for the non-radiative losses, by increasing significantly, the rate at which defect excitons are generated. The occurrence of point defects on the tips becomes all the more plausible, considering that the 1L-WSe$_2$ flake will wrap or tent over the ellipsoid feature. This tenting will subject the portion of the flake close to the tip of the bump to a local tensile strain, which can create point defects in the 1L-WSe$_2$.

	Using equation (1) to define the net PL intensity, the PL enhancement factor $\beta$ is defined as
	\begin{equation}
	\beta = \frac{I_{PL}}{I_{PL}^0}
	\end{equation}
	where I$_{PL}$ is the PL intensity, obtained using equation (1), for the defect exciton dipole placed above the Au ellipsoid, and I$_{PL}^0$ is the PL intensity for the defect exciton dipole above a smooth SiO$_2$/Si substrate. $\beta$ as a function of the aspect ratio $2h/a$ is plotted on the right axes in figures 3a and 3b for 532 and 633 nm excitations respectively. Note that the defect luminescence enhancement can be explained on account of the lightning rod effect, the enhancement factor being lower at 532 nm excitation. Experimentally also, we observe that excitation at 633 nm with 5.8 $\mu$W power exhibits stronger enhancement of the defect luminescence compared to 532 nm excitation with 16 $\mu$W power at certain spots on the flake, as shown in Figure 3d (also in supplementary section S5). Such enhanced emission at 633 nm can be attributed to rough features with a high aspect ratio, for which the electric field enhancement factor is much higher at 633 nm excitation.

We also surmise that the defect peak PL enhancement can include effects other than the lightning rod effect and the quantum yield enhancement. This is because for the PL spectra acquired from many different locations, the enhancement factor of the defect excitons is of a similar order at both 532 and 633 nm excitations (see supplementary figure 1). In fact, the local tensile strain experienced by 1L-WSe$_2$ at the tip of the ellipsoid can also lead to a local funneling of excitons due to a localized decrease in the band gap. The free excitons in the vicinity preferentially populate the defect levels that coincide with the dips or funnels in the local band gap\cite{Kumar2015,branny_deterministic_2017,Palacios-Berraquero2017}, leading to an increase in the defect exciton population.
	
In short, the presence of roughness features on the Au film can significantly strain the 1L-WSe$_2$ film close to the tips. The localized strain in turn creates point defects and also leads to a spatially localized band gap reduction causing exciton funneling, thereby binding free excitons in the close vicinity of the tips to the defects, which leads to an increase in the defect exciton population. Furthermore, there is an enhancement of the excitation laser's electric field intensity at the tips due to the lightning rod effect, which  significantly increases the hot carrier excitation rate. The funneling effect is particularly important for features of low aspect ratio. For features of higher aspect ratio, the lightning rod effect dominates over the band gap funneling effect, which leads to a wavelength sensitive PL enhancement factor $\beta$, as shown in figure 3d. Apart from an increase in the population of the defect excitons, the Purcell effect can also play an important role in overcoming the non radiative losses, thus improving the emission quantum yield of the dipole emitters. 
The excitons, for a given excitation wavelength and power, will be formed randomly over the entire diffraction limited laser spot. The free excitons forming close to the tips of the roughness features are more likely to be bound to the defects due to the band gap funneling. Those which form over the smooth portions do not experience the funneling effect or the lightning rod effect and therefore are more likely to be lost by charge transfer and energy transfer into the Au film. Thus, the free exciton luminescence is limited by the degradation of the PL quantum yield, which is also observed in the electromagnetic simulations, and shown in figures 2c-e for the aspect ratio $h/a = 0$ (red dots).

\subsection{Reduced inhomogenous broadening}

Broadening of spectral lines is a fundamental effect which originates from effects such as finite lifetime in the excited state and interactions of electrons/molecules with each other or the surrounding environment. Two types of spectral broadening are commonly observed : homogenous and inhomogenous. Homogenous broadening arises when the emitting species's lifetime is affected due to its interaction with a uniform environment, and manifests simply as a broadening of the original lifetime limited Lorentzian spectral lineshape. On the other hand, if an emitting species is placed in a fluctuating environment, the random changes in the emission lifetime cause the emission spectrum to acquire a  Gaussian shape, which is termed as "inhomogenous" broadening. For example, the free exciton peaks in 1L-WSe$_2$ can have a strong Gaussian component in the lineshape, physically because of charge traps in the underlying substrate, which are randomly distributed, and play a key role in the screening of the electron and hole Coulomb interaction. In the case of quantum dots, excitons experience a time-dependent Stark effect due to random charging and discharging events in the trap states of the surrounding matrix \cite{Besombes2002,Holmes2015,Seufert2000}. This phenomenon is termed spectral diffusion. Spectral diffusion of the order of a few meV in time scales of several minutes has been observed in the defect luminescence of 1L-WSe$_2$ on SiO$_2$/Si substrate and can also be attributed to the random time dependent Stark effect introduced by the trap states in SiO$_2$ and 1L-WSe$_2$. Often, when an emitter is placed in a complex environment, the linewidth can undergo both homogenous and inhomogenous broadening, and the spectral lineshape exhibits a Voigt profile, which is the convolution of the Lorentzian and Gaussian lineshape.

We investigate any change in the Lorentzian and Gaussian widths of the sharp defect peaks on the two substrates by considering PL spectra taken from different spots of the flake on the two substrates, and fitting the spectra using the Voigt function. The full width at half maximum (FWHM) of the Lorentzian and Gaussian components of the sharp defect peaks are extracted from the fits. Figure 4a shows a scatter plot for the Lorentzian and Gaussian FWHM obtained for the sharp, intense defect peaks at 532 nm excitation and 8 $\mu$W power. The defect peaks for WSe$_2$ on the metal substrate are clustered in a region of narrow Gaussian FWHM. Figure 4b shows the Gaussian and Lorentzian FWHM for the defect peaks obtained at 532 nm with 16 $\mu$W excitation power. Note that the fittings were done for the same number of spectra on both SiO$_2$ and Au, but a significantly higher number of sharp peaks on Au film were observed on each spectrum (see for instance Figure 1c, and also supplementary section S1). A representative PL spectrum on the Au film as a function of the laser excitation power has also been shown in supplementary section S6. At higher power levels, the defect peaks on the SiO$_2$ substrate are not resolvable, because they are obscured by the strongly enhanced charged biexciton peak (see for example figure S2a).
	
As discussed earlier, the Gaussian lineshape of the emission peaks arises from inhomogeneities in the environment of the emitting species.  However, the presence of Au film between 1L-WSe$_2$ and SiO$_2$ can screen the effect of potential fluctuations due to the traps in SiO$_2$. Therefore, the Spectral diffusion of the defect emission lines on the Au film is reduced, which is evidenced by the reducing Gaussian FWHM of the defect peaks as illustrated in figures 4a and 4b.
	
\subsection{Effect of dielectric spacer layer on PL}	
The PL quantum yield of 1L-TMDs is known to be influenced dramatically by the underlying substrate, because of the presence or absence of surface traps and also because of electromagnetic interference effects arising from multiple reflections at the TMD/substrate interface\cite{Jeong2016,Xu2017a,Liu2014}. In this work, the impact of a dielectric spacer between the TMD and metal film is also investigated. Hexagonal Boron Nitride (hBN) of thickness $\sim 5$ nm is chosen as the dielectric spacer. PL from 1L-WSe$_2$ on hBN/SiO$_2$/Si substrate, as well as hBN/Au/SiO$_2$/Si substrate is examined. Spectra for 1l-WSe$_2$/hBN/285 nm SiO$_2$/Si and 1L-WSe$_2$/hBN/50 nm Au/285 nm SiO$_2$/Si  are illustrated in Figure 5 and supplementary section S7. Unlike the spectra in figure 1 and figure S1, it is observed that the entire spectrum enhances in the presence of Au film, and the sharp features are obscured. This is true even at the same excitation powers at which sharp defect peaks were seen in the absence of the dielectric spacer. The enhancement with the introduction of the hBN spacer is not surprising, given that hBN provides a surface free from charge traps, which reduces non-radiative recombination losses, therefore improving the PL quantum yield. Also, being a high band gap material (around 6 eV), hBN acts as a very effective barrier to suppress the loss of free excitons, trions and biexcitons due to charge and energy transfer pathways.Therefore, incorporating a dielectric spacer may not be a good choice as far as observing the sharp quantum emitters with high purity is concerned, as the enhancing background obscures the sharp defect peaks. In fact, as shown in supplementary section S3, the physical contact between the monolayer flake and the rough metal substrate plays an important role in the appearance of the sharp and enhanced defect features. In this case, before the flake is heated, there is known to exist a very small physical separation between the flake and the metal film (equivalent to a very thin dielectric spacer). No sharp peaks are observed in such cases. The sharp and enhanced defect peaks are observed once the substrate is heated, which helps overcome the physical separation between the monolayer flake and the rough metal film.
	
To sum up all the observations, the experimental findings along with the computational analysis demonstrate that it is possible to leverage the excitation electric field enhancement, the Purcell effect and the strain induced exciton funnelling effect, which are spatially localized by nature, to selectively enhance the luminescence of excitons bound to point defects. The metallic film simultaneously quenches the free exciton (and likewise the trion and biexciton) emission. Thus, appropriately patterning the roughness features on metallic substrates can potentially enhance the brightness and purity of quantum emission in 1L-WSe$_2$. For example, the point defects localized to a particular roughness feature can be isolated spatially from other such features by keeping a large enough separation betwen the rough features. Also, in contrast to other works \cite{Johnson2017,Cai2018,Luo,Luo2019,Iff:18} wherein the monolayer film was physically separated from the metal nanostructures by means of thin dielectric spacers, in this report, it is observed that the physical contact between the monolayer flake and the metal film plays an important role in the appearance and the enhancement of the sharp defect features.
	
\renewcommand{\thesection}{\arabic{section}.}
\section{CONCLUSION}
In conclusion, this work points out a simple strategy for improving the defect luminescence, and potentially the quantum emission properties of 1L-WSe$_2$ by utilizing the surface roughness of thin metal films. Transferring 1L-WSe$_2$ directly on a rough metal film leads to a) selective enhancement of defect peaks, b) suppression of free exciton, trion and biexciton luminescence, and c) reduced inhomogeneities through suppression of spectral diffusion. We emphasize that the surface roughness of the metal film and the physical contact of the monolayer flake with the metal film plays a crucial role in this selective spectral enhancement. Rough features add strain to the TMD film and can create point defects at the tips. Furthermore, the strain induced reduction in bandgap further leads to the defect being preferentially populated due to the exciton funneling effect. Moreover, we observed that the PL enhancement also originates from a wavelength dependent strong excitation field enhancement at the tips of these structures, and an increase in the emission quantum yield. On the other hand, the free excitons, excited randomly over a much larger area, arise mostly from smoother portions of the film, and are not subject to the excitation enhancement by the lightning rod effect. Hence, the free exciton PL quenches due to dominance of charge transfer and energy transfer into the Au film. The metal substrate is also beneficial in screening out potential fluctuation in the monolayer, thereby suppressing the spectral diffusion. The incorporation of the dielectric spacer is detrimental from the point of view of observation of the sharp defect features, as the background luminescence arising from the emission of the free many-body quasiparticles enhances significantly. The above findings pave the way for improving the emission properties of quantum emitters in 2D materials in terms of brightness and purity by appropriate engineering of the metal substrate.

\sectionfont{\color{black}}
\subsectionfont{\color{black}}
\section*{SUPPLEMENTARY MATERIAL}
See the supplementary material for more PL spectra on the two substrates, the FDTD simulation details and the PL spectra as a function of the excitation power.

\section*{ACKNOWLEDGMENTS}
This work is supported in part by a grant under Indian Space Research Organization (ISRO), by the grants under Ramanujan Fellowship, Early Career Award, and Nano Mission from the Department of Science and Technology (DST), and by a grant from MHRD, MeitY and DST Nano Mission through NNetRA.

\section*{NOTES}
The authors declare no competing financial interest.

\bibliographystyle{ieeetr}
\bibliography{Manuscript_file_maintext_LATEX_2}

\begin{thebibliography}{10}

\bibitem{Browne}
D.~Browne, S.~Bose, F.~Mintert, and M.~S. Kim, ``{From quantum optics to
  quantum technologies},'' 2017.

\bibitem{Beveratos2002}
A.~Beveratos, A.~Villing, J.~P. Poizat, P.~Grangier, R.~Brouri, and T.~Gacoin,
  ``{Single Photon Quantum Cryptography},'' {\em Physical Review Letters},
  vol.~89, no.~18, p.~187901, 2002.

\bibitem{Giovannetti2011}
V.~Giovannetti, S.~Lloyd, and L.~MacCone, ``{Advances in quantum metrology},''
  2011.

\bibitem{Motes2015}
K.~R. Motes, J.~P. Olson, E.~J. Rabeaux, J.~P. Dowling, S.~J. Olson, and P.~P.
  Rohde, ``{Linear Optical Quantum Metrology with Single Photons: Exploiting
  Spontaneously Generated Entanglement to Beat the Shot-Noise Limit},'' {\em
  Physical Review Letters}, vol.~114, no.~17, 2015.

\bibitem{Aharonovich2016}
I.~Aharonovich, D.~Englund, and M.~Toth, ``{Solid-state single-photon
  emitters},'' {\em Nature Photonics}, vol.~10, no.~10, pp.~631--641, 2016.

\bibitem{Koperski2014a}
M.~Koperski, K.~Nogajewski, A.~Arora, J.~Marcus, P.~Kossacki, and M.~Potemski,
  ``{Single photon emitters in exfoliated WSe$_2$ structures},'' {\em Nature
  Nanotechnology}, vol.~10, no.~6, pp.~503--506, 2014.

\bibitem{Srivastava2014}
A.~Srivastava, M.~Sidler, A.~V. Allain, D.~S. Lembke, A.~Kis, and A.~Imamoglu,
  ``{Optically Active Quantum Dots in Monolayer WSe$_2$},'' {\em Nature
  Nanotechnology}, vol.~10, no.~6, pp.~491--496, 2014.

\bibitem{Chakraborty2015c}
C.~Chakraborty, L.~Kinnischtzke, K.~M. Goodfellow, R.~Beams, and A.~N.
  Vamivakas, ``{Voltage-controlled quantum light from an atomically thin
  semiconductor},'' {\em Nature Nanotechnology}, vol.~10, no.~6, pp.~507--511,
  2015.

\bibitem{Tonndorf2015}
P.~Tonndorf, R.~Schmidt, R.~Schneider, J.~Kern, M.~Buscema, G.~A. Steele,
  A.~Castellanos-Gomez, H.~S.~J. van~der Zant, S.~{Michaelis de Vasconcellos},
  and R.~Bratschitsch, ``{Single-photon emission from localized excitons in an
  atomically thin semiconductor},'' {\em Optica}, vol.~2, no.~4, p.~347, 2015.

\bibitem{He2015b}
Y.~M. He, G.~Clark, J.~R. Schaibley, Y.~He, M.~C. Chen, Y.~J. Wei, X.~Ding,
  Q.~Zhang, W.~Yao, X.~Xu, C.~Y. Lu, and J.~W. Pan, ``{Single quantum emitters
  in monolayer semiconductors},'' {\em Nature Nanotechnology}, vol.~10, no.~6,
  pp.~497--502, 2015.

\bibitem{Palacios-Berraquero2016}
C.~Palacios-Berraquero, M.~Barbone, D.~M. Kara, X.~Chen, I.~Goykhman, D.~Yoon,
  A.~K. Ott, J.~Beitner, K.~Watanabe, T.~Taniguchi, A.~C. Ferrari, and
  M.~Atat{\"{u}}re, ``{Atomically thin quantum light-emitting diodes},'' {\em
  Nature Communications}, vol.~7, pp.~1--6, 2016.

\bibitem{Clark2016}
G.~Clark, J.~R. Schaibley, J.~Ross, T.~Taniguchi, K.~Watanabe, J.~R.
  Hendrickson, S.~Mou, W.~Yao, and X.~Xu, ``{Single defect light-emitting diode
  in a van der Waals heterostructure},'' {\em Nano Letters}, vol.~16, no.~6,
  pp.~3944--3948, 2016.

\bibitem{Zhao2018}
S.~Zhao, J.~Lavie, L.~Rondin, L.~Orcin-Chaix, C.~Diederichs, P.~Roussignol,
  Y.~Chassagneux, C.~Voisin, K.~M{\"{u}}llen, A.~Narita, S.~Campidelli, and
  J.~S. Lauret, ``{Single photon emission from graphene quantum dots at room
  temperature},'' {\em Nature Communications}, vol.~9, no.~1, 2018.

\bibitem{Tran2016}
T.~T. Tran, K.~Bray, M.~J. Ford, M.~Toth, and I.~Aharonovich, ``{Quantum
  emission from hexagonal boron nitride monolayers},'' {\em Nature
  Nanotechnology}, vol.~11, no.~1, pp.~37--41, 2016.

\bibitem{Tran2016c}
T.~T. Tran, C.~Elbadawi, D.~Totonjian, C.~J. Lobo, G.~Grosso, H.~Moon, D.~R.
  Englund, M.~J. Ford, I.~Aharonovich, and M.~Toth, ``{Robust Multicolor Single
  Photon Emission from Point Defects in Hexagonal Boron Nitride},'' {\em ACS
  Nano}, vol.~10, no.~8, pp.~7331--7338, 2016.

\bibitem{Grosso2017}
G.~Grosso, H.~Moon, B.~Lienhard, S.~Ali, D.~K. Efetov, M.~M. Furchi,
  P.~Jarillo-Herrero, M.~J. Ford, I.~Aharonovich, and D.~Englund, ``{Tunable
  and high-purity room temperature single-photon emission from atomic defects
  in hexagonal boron nitride},'' {\em Nature Communications}, vol.~8, no.~1,
  pp.~1--8, 2017.

\bibitem{Bourrellier2016}
R.~Bourrellier, S.~Meuret, A.~Tararan, O.~St{\'{e}}phan, M.~Kociak, L.~H.
  Tizei, and A.~Zobelli, ``{Bright UV single photon emission at point defects
  in h-BN},'' {\em Nano Letters}, vol.~16, no.~7, pp.~4317--4321, 2016.

\bibitem{Ziegler2019}
J.~Ziegler, R.~Klaiss, A.~Blaikie, D.~Miller, V.~R. Horowitz, and B.~J.
  Alem{\'{a}}n, ``{Deterministic Quantum Emitter Formation in Hexagonal Boron
  Nitride via Controlled Edge Creation},'' {\em Nano Letters}, vol.~19, no.~3,
  pp.~2121--2127, 2019.

\bibitem{Zheng2019}
Y.~J. Zheng, Y.~Chen, Y.~L. Huang, P.~K. Gogoi, M.-Y. Li, L.-J. Li, P.~E.
  Trevisanutto, Q.~Wang, S.~J. Pennycook, A.~T.~S. Wee, and S.~Y. Quek,
  ``{Point Defects and Localized Excitons in 2D WSe$_2$},'' {\em ACS Nano},
  2019.

\bibitem{Abdi2018}
M.~Abdi, J.~P. Chou, A.~Gali, and M.~B. Plenio, ``{Color Centers in Hexagonal
  Boron Nitride Monolayers: A Group Theory and Ab Initio Analysis},'' {\em ACS
  Photonics}, vol.~5, no.~5, pp.~1967--1976, 2018.

\bibitem{Tonndorf2017}
P.~Tonndorf, O.~{Del Pozo-Zamudio}, N.~Gruhler, J.~Kern, R.~Schmidt, A.~I.
  Dmitriev, A.~P. Bakhtinov, A.~I. Tartakovskii, W.~Pernice, S.~{Michaelis De
  Vasconcellos}, and R.~Bratschitsch, ``{On-Chip Waveguide Coupling of a
  Layered Semiconductor Single-Photon Source},'' {\em Nano Letters}, vol.~17,
  no.~9, pp.~5446--5451, 2017.

\bibitem{Blauth2018}
M.~Blauth, M.~J{\"{u}}rgensen, G.~Vest, O.~Hartwig, M.~Prechtl, J.~Cerne, J.~J.
  Finley, and M.~Kaniber, ``{Coupling Single Photons from Discrete Quantum
  Emitters in WSe$_2$ to Lithographically Defined Plasmonic Slot Waveguides},''
  {\em Nano Letters}, vol.~18, no.~11, pp.~6812--6819, 2018.

\bibitem{branny_deterministic_2017}
A.~Branny, S.~Kumar, R.~Proux, and B.~D. Gerardot, ``Deterministic
  strain-induced arrays of quantum emitters in a two-dimensional
  semiconductor,'' {\em Nature Communications}, vol.~8, pp.~1--7, May 2017.

\bibitem{Palacios-Berraquero2017}
C.~Palacios-Berraquero, D.~M. Kara, A.~R. Montblanch, M.~Barbone, P.~Latawiec,
  D.~Yoon, A.~K. Ott, M.~Loncar, A.~C. Ferrari, and M.~Atat{\"{u}}re,
  ``{Large-scale quantum-emitter arrays in atomically thin semiconductors},''
  {\em Nature Communications}, vol.~8, no.~15093, 2017.

\bibitem{Johnson2017}
A.~D. Johnson, F.~Cheng, Y.~Tsai, and C.~K. Shih, ``{Giant Enhancement of
  Defect-Bound Exciton Luminescence and Suppression of Band-Edge Luminescence
  in Monolayer WSe$_2$-Ag Plasmonic Hybrid Structures},'' {\em Nano Letters},
  vol.~17, no.~7, pp.~4317--4322, 2017.

\bibitem{Tripathi2018a}
L.~N. Tripathi, O.~Iff, S.~Betzold, {\L}.~Dusanowski, M.~Emmerling, K.~Moon,
  Y.~J. Lee, S.~H. Kwon, S.~H{\"{o}}fling, and C.~Schneider, ``{Spontaneous
  Emission Enhancement in Strain-Induced WSe$_2$ Monolayer-Based Quantum Light
  Sources on Metallic Surfaces},'' {\em ACS Photonics}, vol.~5, pp.~1919--1926,
  may 2018.

\bibitem{Cai2018}
T.~Cai, J.~H. Kim, Z.~Yang, S.~Dutta, S.~Aghaeimeibodi, and E.~Waks,
  ``{Radiative Enhancement of Single Quantum Emitters in WSe$_2$ Monolayers
  Using Site-Controlled Metallic Nanopillars},'' {\em ACS Photonics}, vol.~5,
  pp.~3466--3471, sep 2018.

\bibitem{Luo}
Y.~Luo, G.~D. Shepard, J.~V. Ardelean, D.~A. Rhodes, B.~Kim, K.~Barmak, J.~C.
  Hone, and S.~Strauf, ``{Deterministic coupling of site-controlled quantum
  emitters in monolayer WSe$_2$ to plasmonic nanocavities},'' 2018.

\bibitem{Luo2019}
Y.~Luo, N.~Liu, X.~Li, J.~C. Hone, and S.~Strauf, ``{Single photon emission in
  WSe$_2$ up 160 K by quantum yield control},'' {\em 2D Materials}, vol.~6,
  p.~035017, may 2019.

\bibitem{Iff:18}
O.~Iff, N.~Lundt, S.~Betzold, L.~N. Tripathi, M.~Emmerling, S.~Tongay, Y.~J.
  Lee, S.-H. Kwon, S.~H\"{o}fling, and C.~Schneider, ``Deterministic coupling
  of quantum emitters in wse2 monolayers to plasmonic nanocavities,'' {\em Opt.
  Express}, vol.~26, pp.~25944--25951, Oct 2018.

\bibitem{Castellanos-Gomez2014}
A.~Castellanos-Gomez, M.~Buscema, R.~Molenaar, V.~Singh, L.~Janssen, H.~S. {Van
  Der Zant}, and G.~A. Steele, ``{Deterministic transfer of two-dimensional
  materials by all-dry viscoelastic stamping},'' {\em 2D Materials}, vol.~1,
  p.~011002, apr 2014.

\bibitem{Zhao2013}
W.~Zhao, Z.~Ghorannevis, K.~K. Amara, J.~R. Pang, M.~Toh, X.~Zhang, C.~Kloc,
  P.~H. Tan, and G.~Eda, ``{Lattice dynamics in mono- and few-layer sheets of
  WS$_2$ and WSe$_2$},'' {\em Nanoscale}, vol.~5, no.~20, pp.~9677--9683, 2013.

\bibitem{Huang2016}
J.~Huang, T.~B. Hoang, and M.~H. Mikkelsen, ``{Probing the origin of excitonic
  states in monolayer WSe2},'' {\em Scientific Reports}, vol.~6, mar 2016.

\bibitem{Ugeda2014a}
M.~M. Ugeda, A.~J. Bradley, S.~F. Shi, F.~H. {Da Jornada}, Y.~Zhang, D.~Y. Qiu,
  W.~Ruan, S.~K. Mo, Z.~Hussain, Z.~X. Shen, F.~Wang, S.~G. Louie, and M.~F.
  Crommie, ``{Giant bandgap renormalization and excitonic effects in a
  monolayer transition metal dichalcogenide semiconductor},'' {\em Nature
  Materials}, vol.~13, pp.~1091--1095, dec 2014.

\bibitem{Chernikov2014}
A.~Chernikov, T.~C. Berkelbach, H.~M. Hill, A.~Rigosi, Y.~Li, O.~B. Aslan,
  D.~R. Reichman, M.~S. Hybertsen, and T.~F. Heinz, ``{Exciton binding energy
  and nonhydrogenic Rydberg series in monolayer WS$_2$},'' {\em Physical Review
  Letters}, vol.~113, no.~7, 2014.

\bibitem{Gupta2017a}
G.~Gupta, S.~Kallatt, and K.~Majumdar, ``{Direct observation of giant binding
  energy modulation of exciton complexes in monolayer MoSe$_2$},'' {\em
  Physical Review B}, vol.~96, no.~8, pp.~1--5, 2017.

\bibitem{Schmidt1992}
T.~Schmidt, K.~Lischka, and W.~Zulehner, ``{Excitation-power dependence of the
  near-band-edge photoluminescence of semiconductors},'' {\em Physical Review
  B}, vol.~45, no.~16, pp.~8989--8994, 1992.

\bibitem{Purcell1946}
E.~M. Purcell, H.~C. Torrey, and R.~V. Pound, ``{Resonance absorption by
  nuclear magnetic moments in a solid [7]},'' jan 1946.

\bibitem{Chance1974}
R.~R. Chance, A.~Prock, and R.~Silbey, ``{Lifetime of an emitting molecule near
  a partially reflecting surface},'' {\em The Journal of Chemical Physics},
  vol.~60, no.~7, pp.~2744--2748, 1974.

\bibitem{Chance2007}
R.~R. Chance, A.~Prock, and R.~Silbey, ``{Molecular Fluorescence and Energy
  Transfer Near Interfaces},'' pp.~1--65, John Wiley \& Sons, Ltd, mar 2007.

\bibitem{Chance1975a}
R.~R. Chance, A.~Prock, and R.~Silbey, ``{Comments on the classical theory of
  energy transfer},'' {\em The Journal of Chemical Physics}, vol.~62,
  pp.~2245--2253, 1975.

\bibitem{Liao1982}
P.~F. Liao and A.~Wokaun, ``{Lightning rod effect in surface enhanced Raman
  scattering},'' {\em The Journal of Chemical Physics}, vol.~76, no.~1,
  pp.~751--752, 1982.

\bibitem{Bharadwaj2009}
P.~Bharadwaj, B.~Deutsch, and L.~Novotny, ``{Optical Antennas},'' {\em Advances
  in Optics and Photonics}, vol.~1, no.~3, p.~438, 2009.

\bibitem{Novotny2012}
L.~Novotny, B.~Hecht, L.~Novotny, and B.~Hecht, ``{Propagation and focusing of
  optical fields},'' in {\em Principles of Nano-Optics}, pp.~45--85, Cambridge:
  Cambridge University Press, 2012.

\bibitem{Wang2016}
Z.~Wang, Z.~Dong, Y.~Gu, Y.~H. Chang, L.~Zhang, L.~J. Li, W.~Zhao, G.~Eda,
  W.~Zhang, G.~Grinblat, S.~A. Maier, J.~K. Yang, C.~W. Qiu, and A.~T. Wee,
  ``{Giant photoluminescence enhancement in tungsten-diselenide-gold plasmonic
  hybrid structures},'' {\em Nature Communications}, vol.~7, p.~11283, dec
  2016.

\bibitem{Kumar2015}
S.~Kumar, A.~Kaczmarczyk, and B.~D. Gerardot, ``{Strain-Induced Spatial and
  Spectral Isolation of Quantum Emitters in Mono- and Bilayer WSe$_2$},'' {\em
  Nano Letters}, vol.~15, no.~11, pp.~7567--7573, 2015.

\bibitem{Besombes2002}
L.~Besombes, K.~Kheng, L.~Marsal, and H.~Mariette, ``{Few-particle effects in
  single CdTe quantum dots},'' {\em Physical Review B}, vol.~65, no.~12,
  p.~121314, 2002.

\bibitem{Holmes2015}
M.~Holmes, S.~Kako, K.~Choi, M.~Arita, and Y.~Arakawa, ``{Spectral diffusion
  and its influence on the emission linewidths of site-controlled GaN nanowire
  quantum dots},'' {\em Physical Review B - Condensed Matter and Materials
  Physics}, vol.~92, no.~11, pp.~1--7, 2015.

\bibitem{Seufert2000}
J.~Seufert, R.~Weigand, G.~Bacher, T.~K{\"{u}}mmell, A.~Forchel, K.~Leonardi,
  and D.~Hommel, ``{Spectral diffusion of the exciton transition in a single
  self-organized quantum dot},'' {\em Applied Physics Letters}, vol.~76,
  no.~14, pp.~1872--1874, 2000.

\bibitem{Jeong2016}
H.~Y. Jeong, U.~J. Kim, H.~Kim, G.~H. Han, H.~Lee, M.~S. Kim, Y.~Jin, T.~H. Ly,
  S.~Y. Lee, Y.~G. Roh, W.~J. Joo, S.~W. Hwang, Y.~Park, and Y.~H. Lee,
  ``{Optical Gain in MoS$_2$ via Coupling with Nanostructured Substrate:
  Fabry-Perot Interference and Plasmonic Excitation},'' {\em ACS Nano},
  vol.~10, pp.~8192--8198, sep 2016.

\bibitem{Xu2017a}
H.~Xu, ``{Enhanced light-matter interaction of a MoS$_2$ monolayer with a gold
  mirror layer},'' {\em RSC Advances}, vol.~7, no.~37, pp.~23109--23113, 2017.

\bibitem{Liu2014}
J.~T. Liu, T.~B. Wang, X.~J. Li, and N.~H. Liu, ``{Enhanced absorption of
  monolayer MoS$_2$ with resonant back reflector},'' {\em Journal of Applied
  Physics}, vol.~115, p.~193511, may 2014.

\end{thebibliography}


\begin{thebibliography}{1}

\bibitem{Huang2016}
J.~Huang, T.~B. Hoang, and M.~H. Mikkelsen, ``{Probing the origin of excitonic
  states in monolayer WSe2},'' {\em Scientific Reports}, vol.~6, mar 2016.

\bibitem{Li2018}
Z.~Li, T.~Wang, Z.~Lu, C.~Jin, Y.~Chen, Y.~Meng, Z.~Lian, T.~Taniguchi,
  K.~Watanabe, S.~Zhang, D.~Smirnov, and S.~F. Shi, ``{Revealing the biexciton
  and trion-exciton complexes in BN encapsulated WSe2},'' {\em Nature
  Communications}, vol.~9, no.~1, 2018.

\bibitem{PhysRevB.6.4370}
P.~B. Johnson and R.~W. Christy, ``Optical constants of the noble metals,''
  {\em Phys. Rev. B}, vol.~6, pp.~4370--4379, Dec 1972.

\bibitem{Kaminski2007a}
F.~Kaminski, V.~Sandoghdar, and M.~Agio, ``{Finite-difference time-domain
  modeling of decay rates in the near field of metal nanostructures},'' {\em
  Journal of Computational and Theoretical Nanoscience}, vol.~4, pp.~635--643,
  may 2007.

\bibitem{Xu}
Y.~Xu, R.~K. Lee, and A.~Yariv, ``{Quantum analysis and the classical analysis
  of spontaneous emission in a microcavity},'' {\em Physical Review A},
  vol.~61, no.~3, p.~033807, 2000.

\bibitem{Bharadwaj2009}
P.~Bharadwaj, B.~Deutsch, and L.~Novotny, ``{Optical Antennas},'' {\em Advances
  in Optics and Photonics}, vol.~1, no.~3, p.~438, 2009.

\end{thebibliography}
\pagebreak

\begin{figure*}[!hbt]
	\includegraphics[scale=0.5] {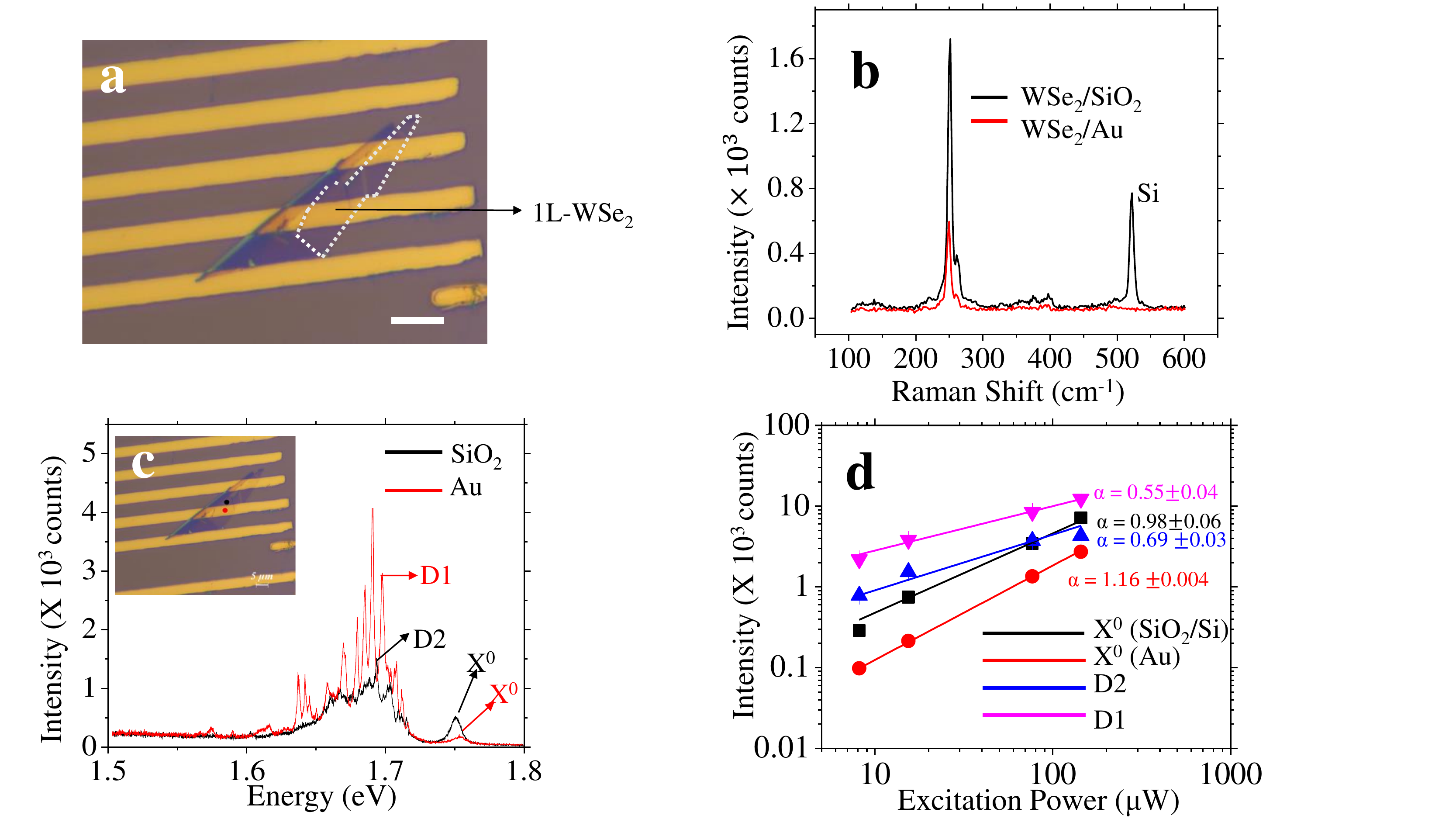}
	\caption{\textbf{Optical characterization of the 1L-WSe$_2$ flake.} (a) Optical micrograph of the 1L-WSe$_2$ flake. The white dashed lines highlight the edges of the monolayer flake. The scale bar at the bottom right is 10 $\mu$m. (b) Raman spectra at 532 nm excitation and 10 seconds acquisition for the flake on SiO$_2$/Si (black trace) and on the Au film (red trace). (c) Representative PL spectra for the 1L-WSe$_2$ flake on SiO$_2$ (black trace) and on Au film (red trace) with the corresponding locations on the flake labelled in the inset. The spectra were acquired at 532 nm excitation, 8 $\mu$W power over an acquisition time of 30 seconds. X$^0$ stands for the free exciton. (d) The power dependence of the PL intensity for the exciton peaks as well as the peaks D1 an D2 labelled in figure 1c, fit using the relation $\log(I)$ = $\alpha$$\log(P) + c$. The coefficients $\alpha$ have also been shown for the corresponding emission peaks.}\label{fig:F1}
\end{figure*}

\pagebreak

\begin{figure*}[!hbt]
	\includegraphics[scale=0.5] {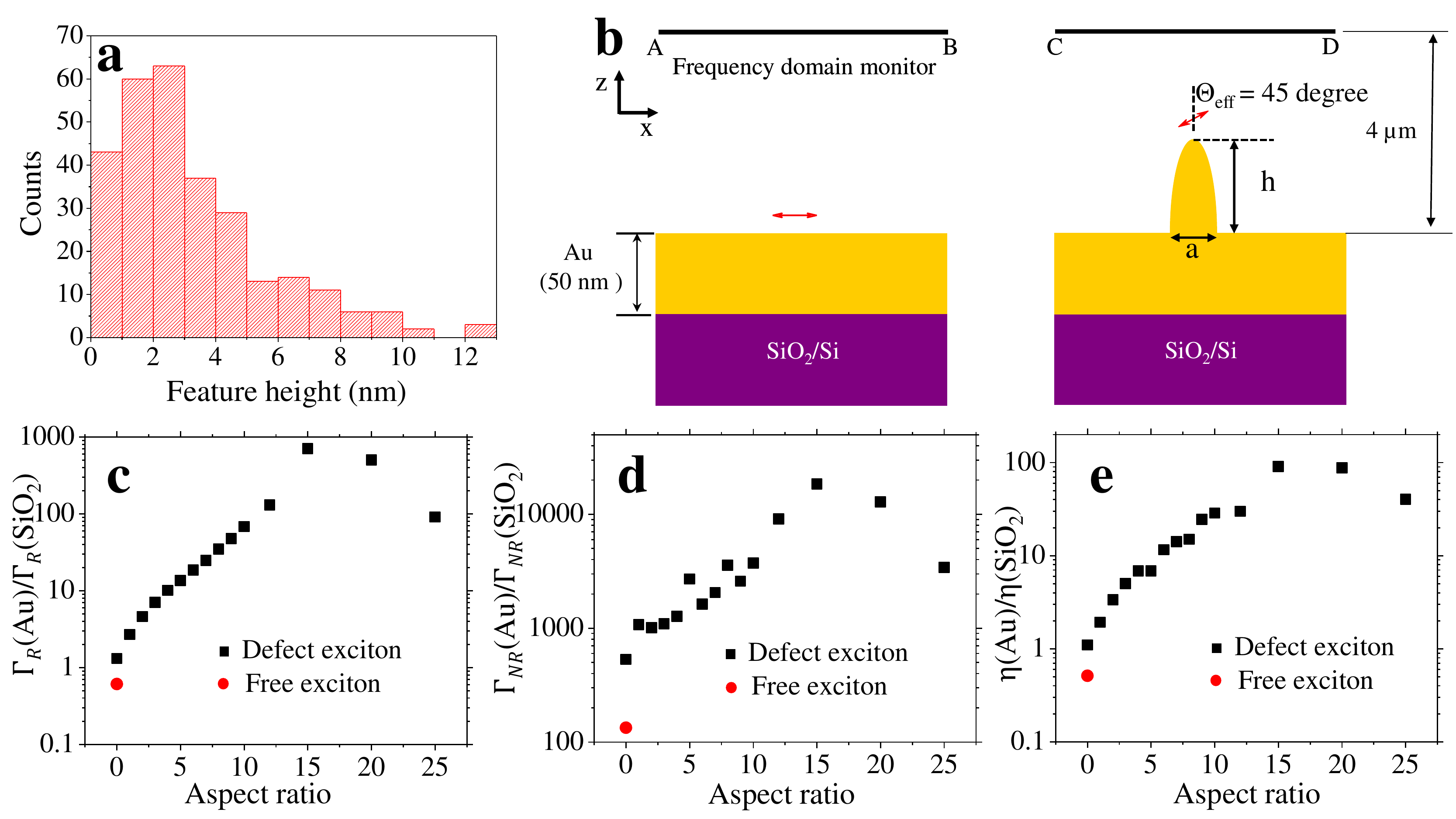}
	\caption{\textbf{Exciton emission quantum yield analysis} (a) Histogram for the height of the rough features on the portions of the Au film supporting the 1L-WSe$_2$, obtained by AFM. (b) Simulation schematic for the Quantum yield simulation for the free exciton (left) and the defect exciton (right). The free exciton is modeled as a dipole source 4 {\AA} above a smooth Au film and is radiating at 710 nm (1.75 eV). The defect exciton is modeled as a dipole emitter at 753 nm (1.65 eV) 4 {\AA} above the tip of an ellipsoid of diameter a = 2 nm and varying height h, inclined at an angle $\theta_{eff}$ = 45 degree, as discussed in the main text. (c) The normalized radiative decay rate, (d) normalized non-radiative decay rate and (e) the emission quantum yield enhancement factor as a function of the ellipsoid aspect ratio ($2h/a$). The red dots represent the respective values for the free exciton dipole. The enhancement factors were normalized with respect to the same dipole emitter 4 {\AA} above a smooth SiO$_2$/Si film.}\label{fig:F2}
\end{figure*}

\pagebreak

\begin{figure*}[!hbt]
	\includegraphics[scale=0.5] {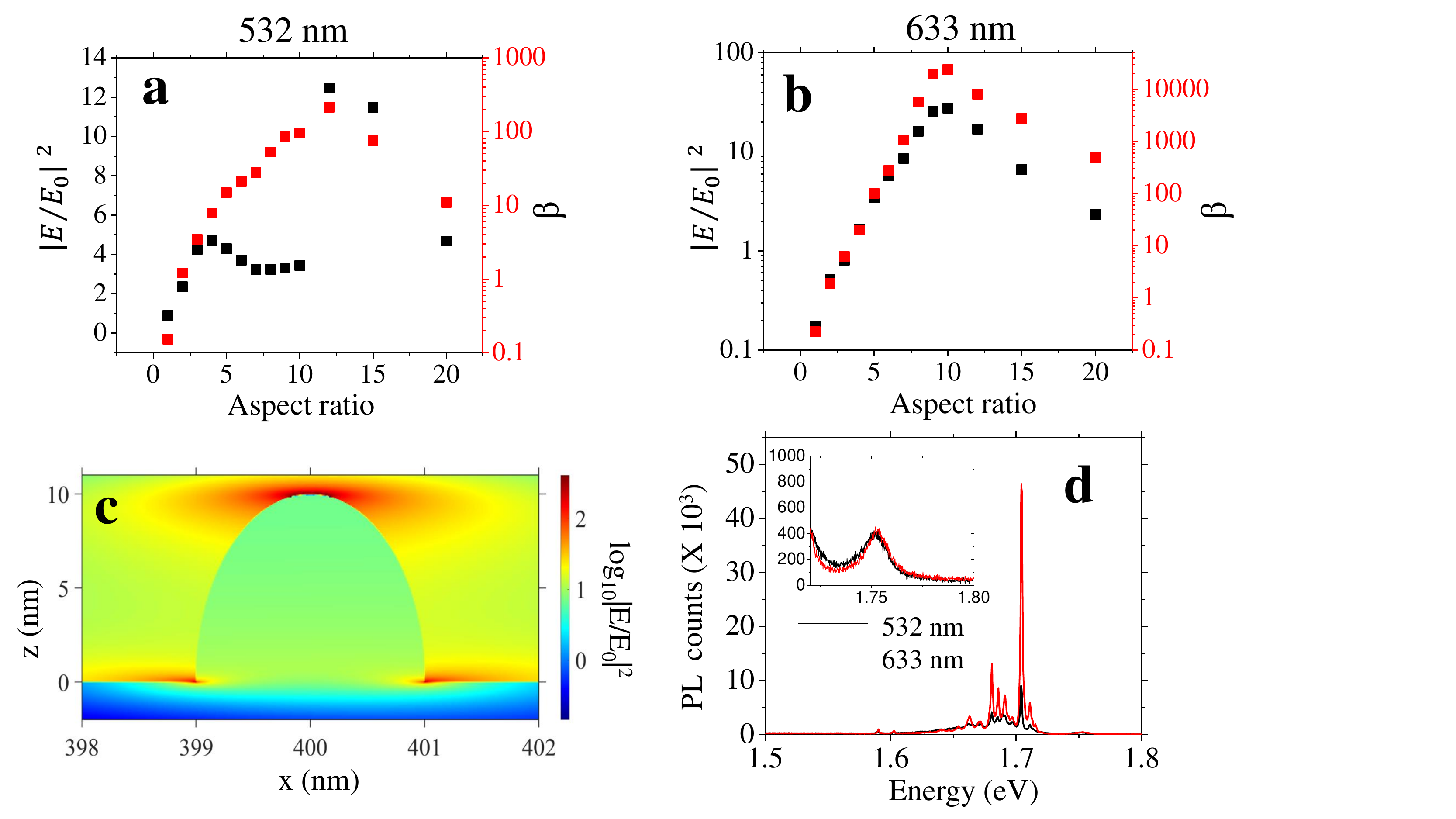}
	\caption{\textbf{Localized excitation field enhancement due to surface roughness} The excitation electric field intensity enhancement factor 4 \AA above the tip of the Au ellipsoid (black axis) and the total PL enhancement factor $\beta$ (red axis) as a function of the ellipsoid aspect ratio are plotted, for (a) 532 nm and (b) 633 nm excitations. (c) Representative colour map of the Electric field intensity enhancement at 633 nm excitation for Au ellipsoid of height 10 nm and diameter 2 nm. (d) Comparison of the PL from a particular spot on 1L-WSe$_2$ on the Au film at 532 nm, 16 $\mu$W excitation and 633 nm, 5.8 $\mu$W excitation, displaying higher enhancement of the defect peak luminescence at 633 nm excitation. Inset to the figure shows the comparison of the free exciton peak at the two excitation wavelengths.
	}\label{fig:F3}
\end{figure*}
\pagebreak

\begin{figure*}[!hbt]
	\includegraphics[scale=0.55] {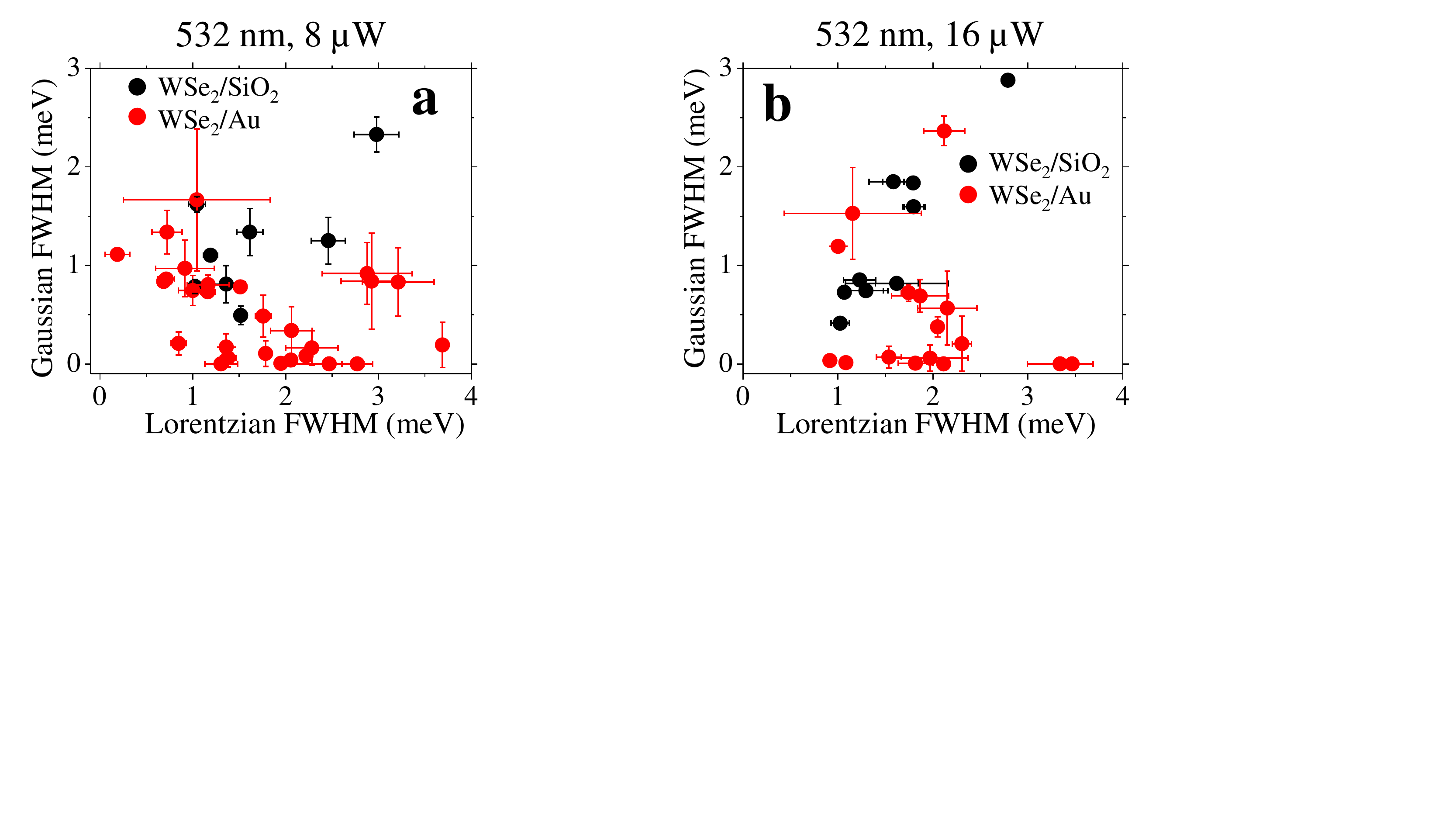}
	\caption{\textbf{Linewidth of the defect emission peaks.} Scatter plots of the Gaussian and Lorentzian FWHM for the sharp defect peaks at (a) 532 nm, 8 $\mu$W power excitation and (b) 532 nm, 16 $\mu$W power excitation. All spectra were acquired for 30 seconds. The defect peaks for 1L-WSe$_2$ on Au film are mostly clustered in regions of lower Gaussian linewidth. The broadening of the Lorentzian FWHM is attributed to the additional charge transfer and energy transfer pathways in the presence of the Au film.}\label{fig:F4}
	
\end{figure*}
\pagebreak

\begin{figure*}[!hbt]
	\includegraphics[scale=0.55] {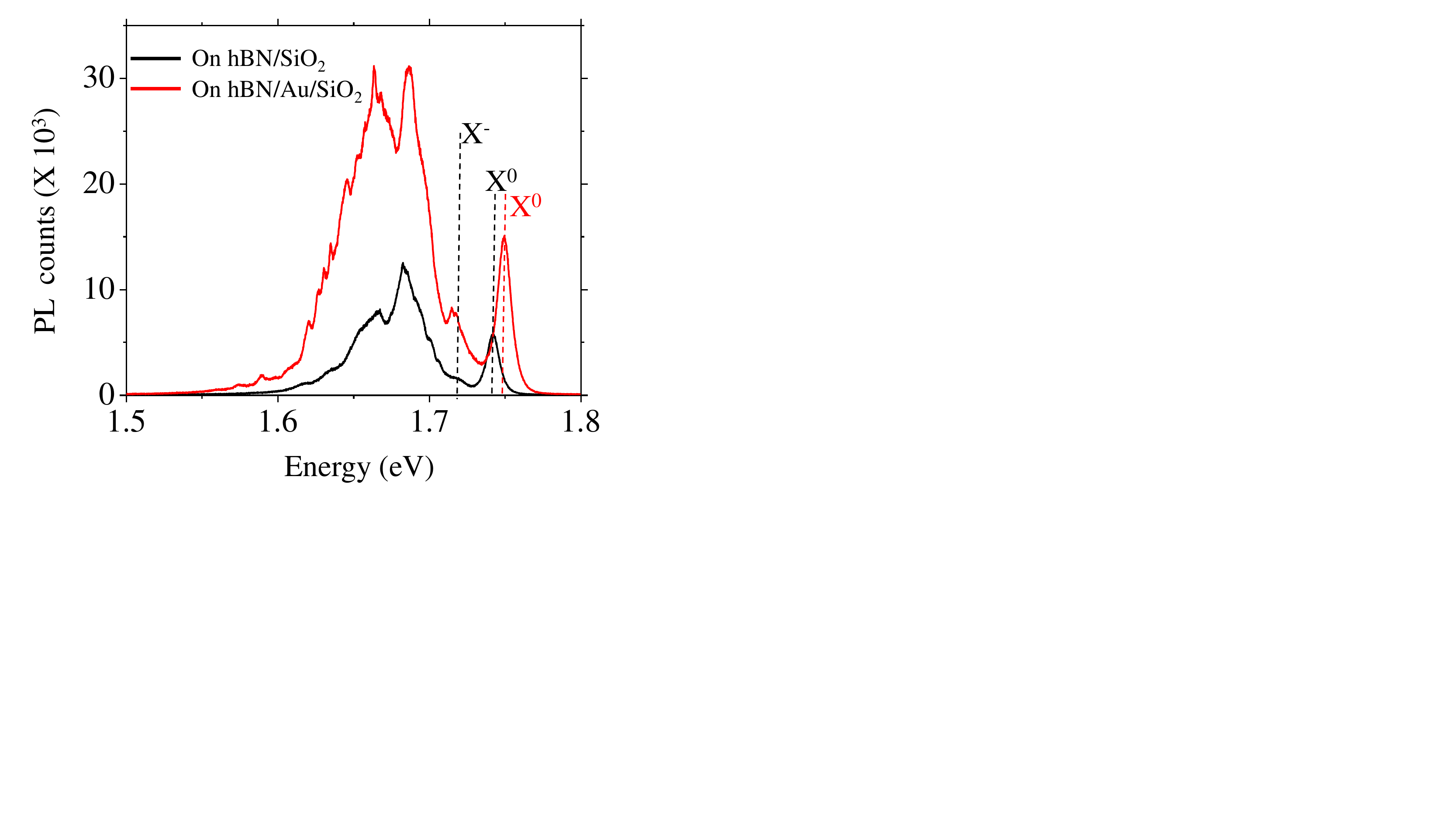}
	\caption{\textbf{PL spectra for 1L-WSe$_2$ on hBN/SiO$_2$/Si (black) and hBN/Au/SiO$_2$/Si (red).} The excitation laser is fixed at 532 nm, 16 $\mu$W excitation. The PL spectra are acquired for 30 seconds. The hBN thickness is close to 5 nm. Sandwiching hBN between the Au film and 1L-WSe$_2$ enhances the free exciton (labelled as X$^0$), trion (labelled as X$^-$) and biexciton emission along with the defect luminescence, thus obscuring the sharp defect peaks and reducing the purity of quantum light emission from 1L-WSe$_2$.}\label{fig:F5}
	
\end{figure*}
\pagebreak
	
\end{document}


\maketitle
	
	\renewcommand*{\thesection}{S\arabic{section}}
	\renewcommand{\thefigure}{S\arabic{figure}} 

	\section{PL spectra for 1L-WSe$_2$/285 nm SiO$_2$/Si and 1L-WSe$_2$/50 nm Au/285 nm SiO$_2$/Si}
		 Figure S1 shows the PL spectra of 1L-WSe$_2$ flake on SiO$_2$/Si substrates (black traces) and on Au film (red traces) at (a),(b)532 nm, 16 $\mu$W excitation and (c),(d) 633 nm, 5.8 $\mu$W excitation. The window to the right of each spectrum compares the free exciton peaks on the two substrates. The spectra were acquired for a total of 30 seconds integration time.

\begin{figure*}[!hbt]
	\includegraphics[scale=0.5] {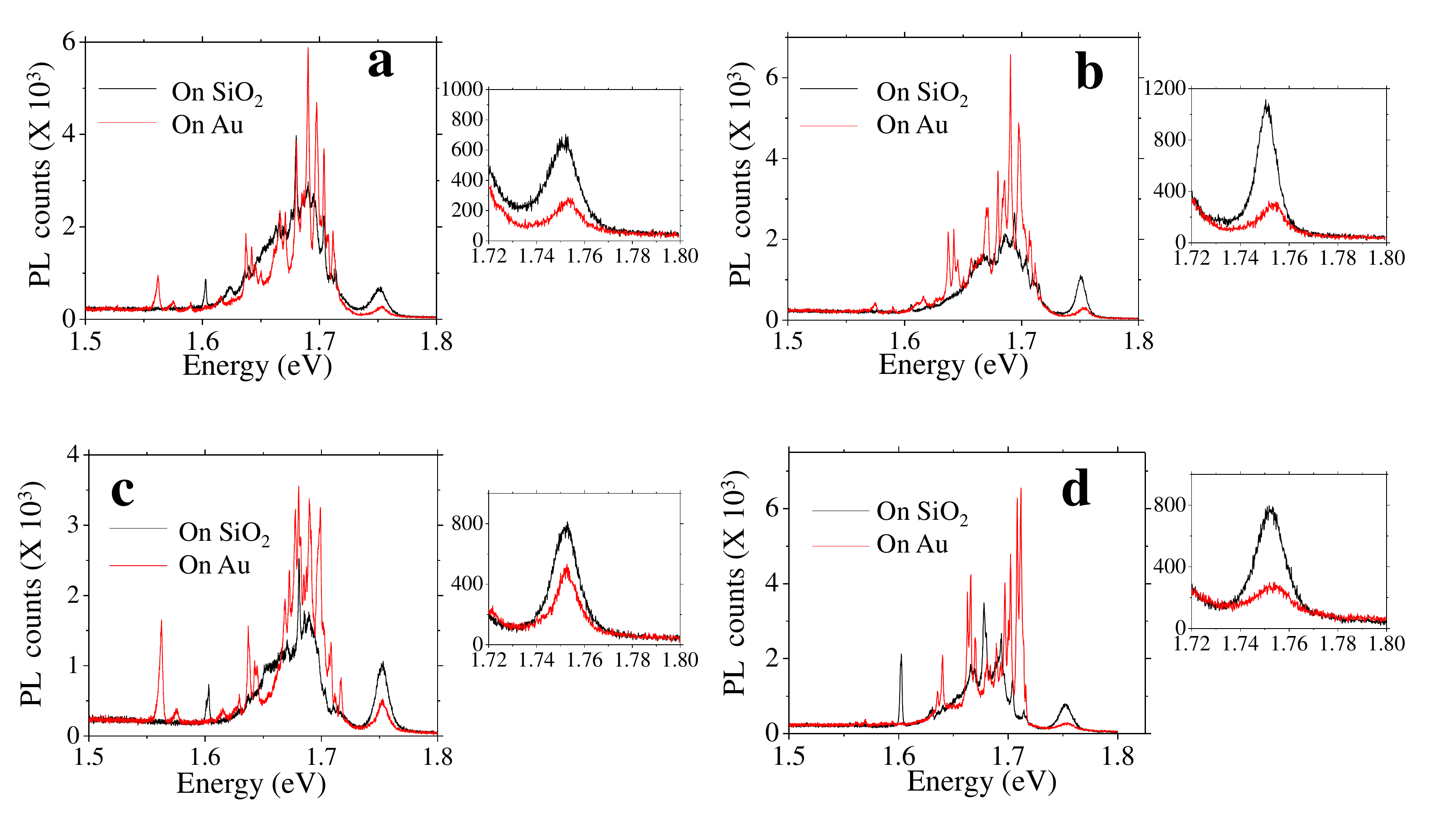}
	\caption{\textbf{PL spectra for 1L-WSe$_2$ on SiO$_2$/Si (black trace) and Au/SiO$_2$/Si (red trace).} Spectra acquired at (a),(b) 532 nm, 16 $\mu$W excitation and (c),(d) 633 nm, 5.8 $\mu$W excitation. A comparison of the free exciton peaks is also shown to the right of each spectrum. All spectra were acquired over a 30 seconds acquisition time window.} \label{fig:S1}
\end{figure*}

\pagebreak
	
	\section{Power dependent PL spectra}
	Figures S2a and S2b plot the PL spectra from the same points as those chosen in figure 1c as a function of the laser excitation power, on the SiO$_2$ and the Au substrate respectively. The laser excitation was fixed at 532 nm continuous wave and the spectra were acquired over a 30 seconds acquisition interval. At higher powers, on both the substrates, a shoulder peak around 1.71 eV (labelled X$^-$ in figures S2a and S2b is also observed). This peak position has been reported in the literature to correspond to the negatively charged trion peak \cite{Huang2016}. The WSe$_2$ flake on SiO$_2$/Si flake also displays a feature (labelled as XX$^-$ in figure S2a) centred at 1.68 eV, which is seen to increase super-linearly with respect to power. This super-linear increase points towards a bi-excitonic state, more specifically the exciton-trion state, as reported previously \cite{Li2018}. Note that on the metal substrate, the trion and biexcitonic luminescence is also seen to quench, just like the excitonic state. It is also noted that at higher powers, the sharp defect features on the WSe$_2$ flake on metal substrates become less prominent because of the increase in the background luminescence.   
	
		  \begin{figure*}[!hbt]
		\includegraphics[scale=0.5] {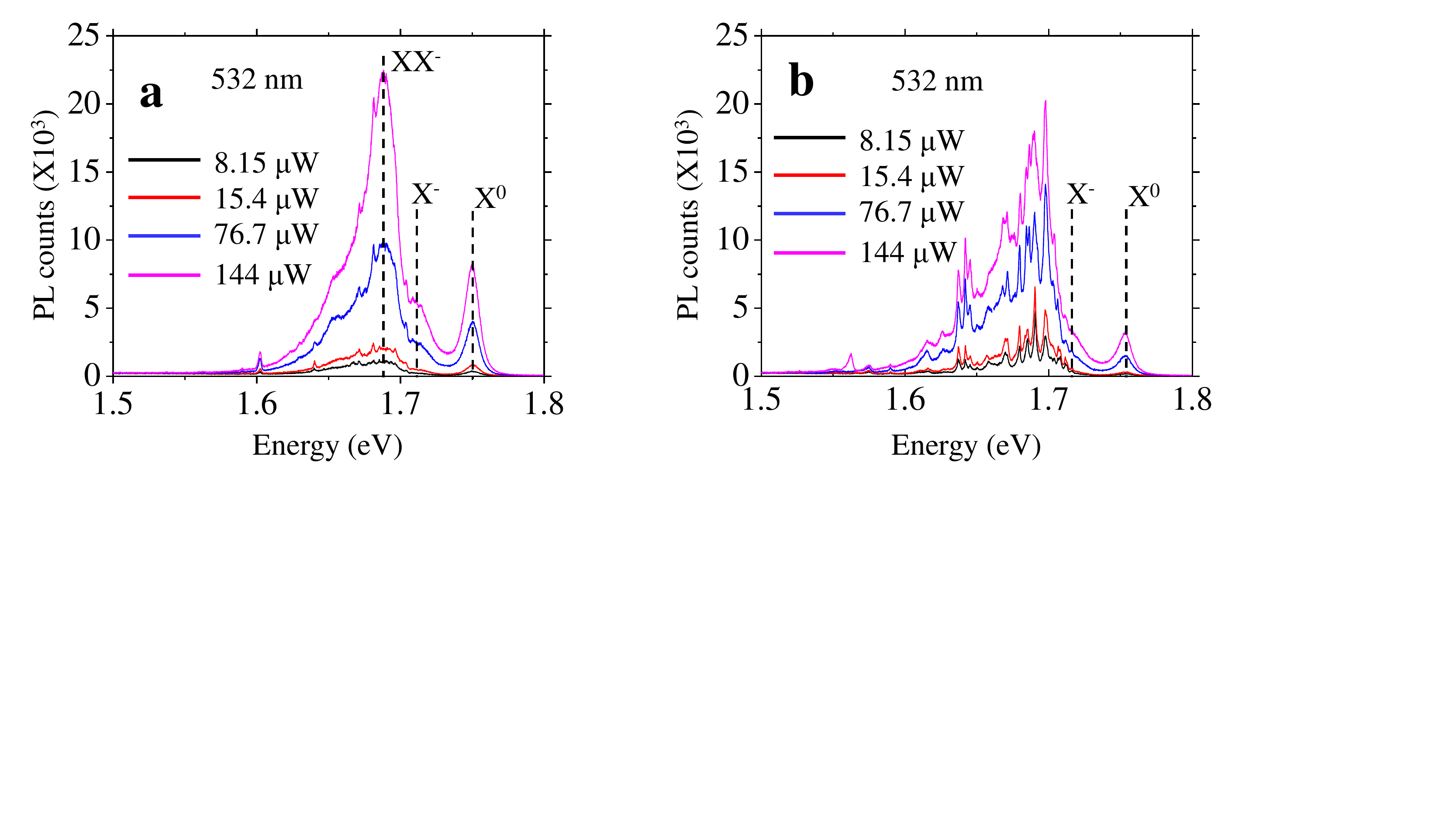}
	\vspace{-1.5in}
		 	\caption{\textbf{Excitation power dependence of the PL spectra}. The PL spectra plotted as a function of the excitation power for 1L-WSe$_2$ on (a)SiO$_2$/Si substrate and (b) Au film. The excitation wavelength was fixed at 532 nm and the acquistion time was 30 seconds. The trion (X$^-$) and charged biexcitonic (XX$-$) luminescence features are prominent at higher powers, and are observed to quench on the metal film.} \label{fig:S2}
	 \end{figure*}
 
  \pagebreak

  \section{The role of heating in the appearance of sharp defect peaks}
Figure S3a shows a 1L-WSe$_2$ flake transferred partially on the SiO$_2$/Si substrate and partially on a separate 50 nm thick Au film, covered partially by around 15 nm thick hBN. As a first experiment, Photoluminescence was acquired at 5 Kelvin immediately after transferring the monolayer flake on the Au film, using the same dry transfer technique as employed in the main text. In a second experiment, the substrate was heated on a hot plate at a temperature of 75 degree celsius for 2 minutes under ambient conditions, and PL acquired at the same temperature (5 Kelvin). The annealing reduces the physical separation between the WSe$_2$ layer and the metal substrate, as evidenced by the suppression and slight shift of the neutral exciton peak (at location 2). However, we do not observe any significant change in the features of the neutral free exciton and the defect peaks on hBN (location 1) after the annealing step. The comparison of the data before and after annealing allows us to probe the same location (and thus the same defect sites) as a function of the distance from the Au substrate, clearly demonstrating the role of the metal. Figure S3d depicts the histogram of the roughness feature heights of the portion of the metal film supporting the monolayer flake, which confirms the roughness of the Au film surface. The monolayer flake in contact with the rough Au film is subjected to the Purcell effect and the lightning rod effect as discussed in the main text, which explains the selective enhancement of the defect peaks observed for the monolayer flake on the Au film.

 \begin{figure*}[!hbt]
	\includegraphics[scale=0.5] {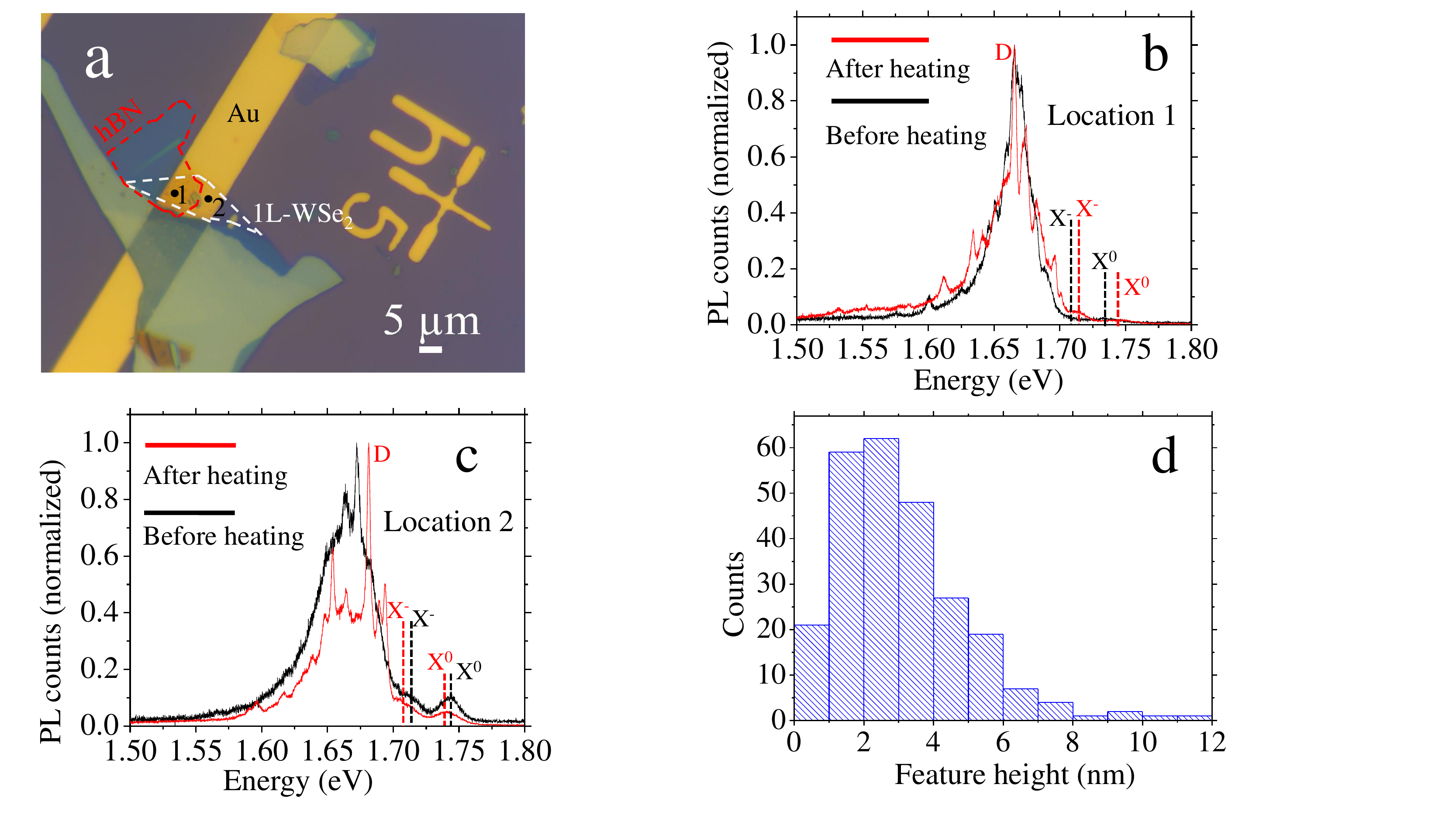}
	\caption{(a) Optical image of the stack used for measurement. Spots 1 and 2 denote the locations chosen for the PL spectra depicted in figures S3b and S3c respectively (b) PL spectrum taken before heating (black trace) and after heating (red trace) from location 1 in figure S3a. (c) PL spectrum taken before heating (black trace) and after heating (red trace) from location 2 in figure S3a. All PL spectra acquired at 532 nm excitation, 2.67 $\mu$W power at 5 Kelvin temperature and 30 seconds accumulation time. X$^0$ denotes the neutral exciton peak, X$^-$ denotes the charged exciton (trion) peak and D denotes the sharp defect peak  (d) Histogram of the roughness feature heights obtained from AFM on the portion of the Au film covered by the monolayer flake.} \label{fig:S3}
\end{figure*}

\pagebreak

	 \section{FDTD Simulations}

	 \subsection{Quantum Yield simulations}
	
	 For the Quantum yield simulations, a 2D cross-sectional geometry (along the XZ-plane) was considered for the cases of 1L-WSe$_2$ on Au/SiO$_2$/Si and on SiO$_2$/Si substrates, as illustrated in figure 2b of the main text.
	 A point dipole source is used to simulate the spontaneous emission of the free and defect excitons. The free exciton is modeled as a dipole emitter oriented along the X-direction (in the plane of the 1L-WSe$_2$) with emission fixed at 710 nm (1.75 eV), and is placed 4 {\AA} above a smooth Au film 50 nm thick. For the defect peak emission, the dipole emission is fixed at 753 nm (~1.65 eV).The defect excitons are chosen to lie 4 {\AA} above the tip of a Au ellipsoid on a 50 nm thick Au film. To consider the effect of all possible dipole orientations, the defect exciton dipole is inclined at an angle of 45 degree with respect to the major axis of the ellipsoid (see main text). The base diameter ($a$) of the ellipsoid is fixed at 2 nm and the height, denoted by the variable h in figure 2b of the main text, is varied. The Johnson and Christy model\cite{PhysRevB.6.4370} is used to define the optical constants of the Au film and the ellipsoid. The optical constants of SiO$_2$ and Si are obtained from the Palik model. A mesh override region is defined about the ellipsoid feature of size 20 nm along the x direction and 30 nm along the z direction. The mesh spacing in this override region is kept at 0.05 nm along both x and z directions. A 1D frequency domain monitor of length 8 $\mu$m (along the X axis, denoted by lines AB and CD in figure 2b of the main text) is placed 4 $\mu$m above the Au film to record the power radiated by the dipole (in Watt/metre), which is evaluated as the integral of the Poynting vector crossing the line. Though the dipole radiates in all directions, the upper monitor is sufficient, considering that the microscope objective in the experimental setup is vertically above the sample. The power crossing the line monitor is normalized with respect to the power emitted by the dipole in free space to quantify the normalized radiative decay rate\cite{Kaminski2007a,Xu}. The ratio of the normalized radiative decay rates for the dipole close to Au to that of the same dipole close to smooth SiO$_2$/Si substrate (in the absence of the Au film) is quantified as the decay rate enhancement. The non-radiative decay rate is quantified by absorption losses in the underlying substrate. The power absorbed over all the substrate layers is added up to obtain the total power absorbed, and is normalized with respect to power emitted by the dipole in free space\cite{Kaminski2007a} to obtain the normalized non-radiative decay rate. The ratio of the normalized non-radiative decay rates on Au to that on SiO$_2$/Si substrate is quantified as the non-radiative decay rate enhancement. Both x and z boundaries are set as perfectly matched layers (PML boundary conditions). The PL Quantum yield was evaluated using the relation
	
	 \begin{equation}	
	 QY = \frac{P_{rad}}{P_{rad}+P_{lost}}
	 \end{equation}
	
	 where P$_{rad}$ denotes the power radiated by the dipole into the far field (total power crossing the line monitor) and P$_{lost}$ denotes the total power loss. The total power loss includes both the intrinsic losses in the TMD (owing to non-radiative recombination losses like phonon scattering) as well as absorption losses in the substrate. To separate out the intrinsic losses from the substrate losses, the radiated and lost powers are normalized with respect to the power emitted by the dipole in free space (P$_0$). Then, one obtains \cite{Bharadwaj2009}
	
	 \begin{equation}	
	 QY = \frac{(P_{rad}/P_0)}{(P_{rad}/P_0)+(P_{substrate}/P_0)+(1-\eta_i)/\eta_i}
	 \end{equation}
	
	 where $\eta_i$ refers to the intrinsic Quantum Yield of the material, P$_{substrate}$ refers to the absorption losses in the substrate. For the purpose of this simulation $\eta_i$ is chosen as 0.001.

	 \subsection{Electric field enhancement simulations}
	
	 The roughness feature on the Au film is modeled as an ellipsoid of diameter $a = 2 nm$ and height ranging from 0 nm to 25 nm (variable h in figs S2a and S2b). The simulations are run over the 2D cross-section geometry (along the XZ-plane). The excitation laser beam is modeled as a Gaussian beam propagating in the negative Z direction polarized along the X direction. The Gaussian beam is modeled as emanating from a thin lens of numerical aperture NA 0.5, in accordance with the experimental setup used. It is also assumed that the entire back plane of the lens is illumined. Perfectly Matched Layer (PML) boundary conditions are imposed along the simulation boundaries. A 2D frequency domain monitor around the ellipsoid (oriented parallel to the XZ-plane) is used to obtain the colour map of the electric field intensity. A mesh override region of 2 nm extent along the x-direction and 30 nm extent along z direction is defined around the ellipsoid, with a spacing of 0.05 nm along x direction and 0.1 nm along z direction. The optical constants of Au are extracted from the Johnson and Christy model, while the optical constants of SiO$_2$ and Si are obtained using the Palik model. A representative colour map at 633 nm excitation for an ellipsoid 7 nm tall is shown in figure S2c. As pointed out in the main text, a significant enhancement of the excitation electric field intensity can be observed close to the ellipsoid tip. On account of the tensile strain experienced by the TMD flake at the tip of the ellipsoid, point defects can be created close to the tip of the ellipsoid. The enhanced excitation laser field intensity as well as the strain-induced exciton funneling effect contribute to an increase in the number of defect excitons created.
	
	 	\begin{figure*}[!hbt]
	 	\includegraphics[scale=0.5] {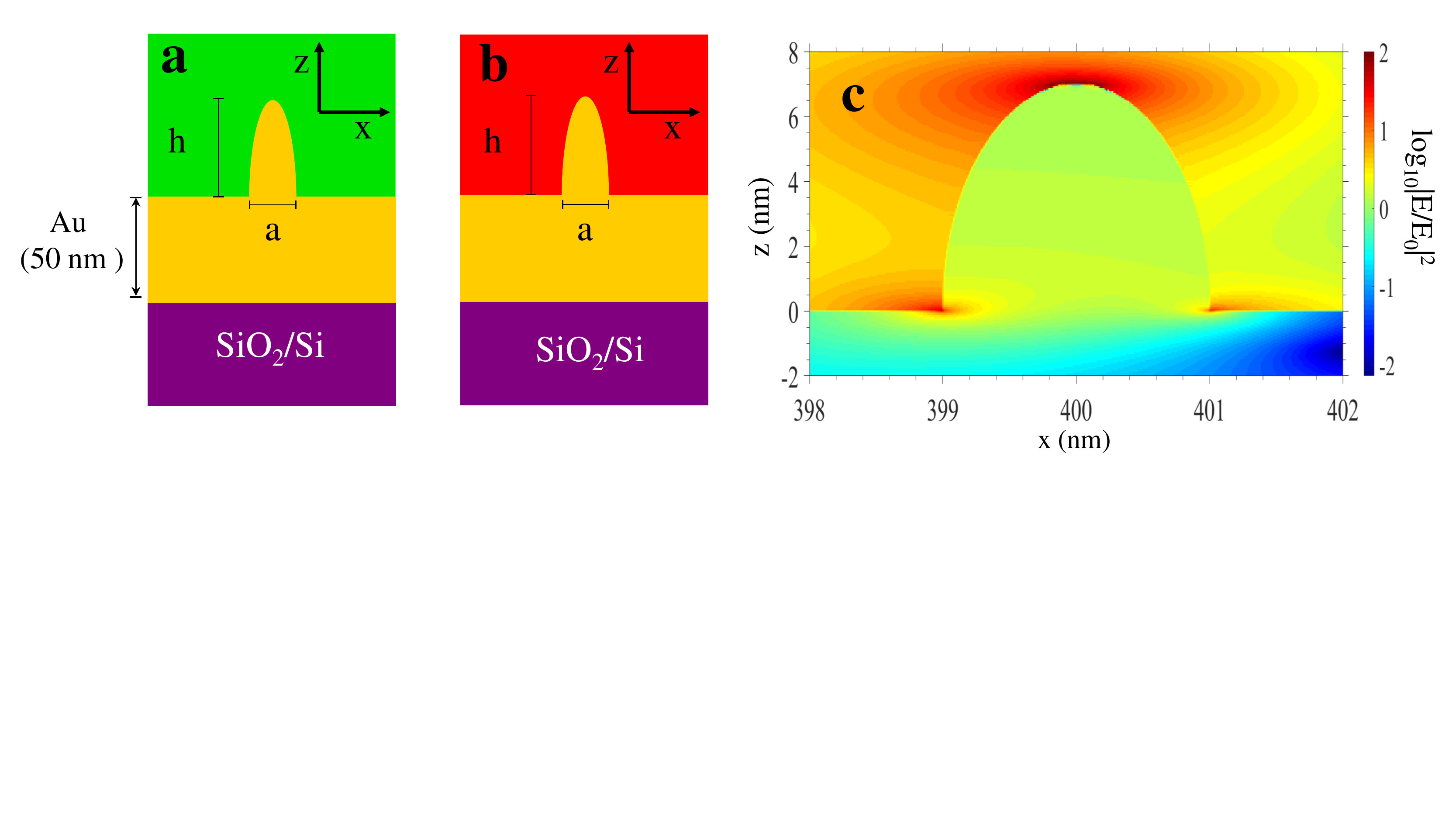}
	 	 \vspace{-1.5in}
	 	\caption{\textbf{Electric field enhancement } (a) Rough feature on Au film, at 532 nm excitation (left) and (b) 633 nm excitation (right). (c) Representative colour plot for the field enhancement at 633 nm excitation for ellipsoid height of 7 nm. Note that the field enhancement is concentrated at the tip of the ellipsoid, where it is likely to observe defects due to the additional strain introduced when the flake wraps on the ellipsoid feature. $x = 0$ denotes the centre of the Gaussian beam cross section.} \label{fig:S4}
	 \end{figure*}

	 \pagebreak
	
	
	

	 \section{Excitation wavelength dependent enhancement of the defect peaks}
	
	 Figures S5 shows a spectrum taken from a particular diffraction limited spot on the 1L-WSe$_2$ flake on Au film at 532 nm excitation, 16$\mu$W power (black trace) and 633 nm excitation, 5.8 $\mu$W power (red trace). All spectra were acquired over an integration time of 30 seconds. The defect peaks are more enhanced at 633 nm compared to 532 nm excitation, which can be attributed to the lightning rod effect, which is more pronounced at 633 nm excitation. Such enhancements are possibly arising from roughness features of higher aspect ratio. The inset compares the free exciton peaks at 532 nm and 633 nm excitation.
	
	 \begin{figure*}[!hbt]
	 	\includegraphics[scale=0.5] {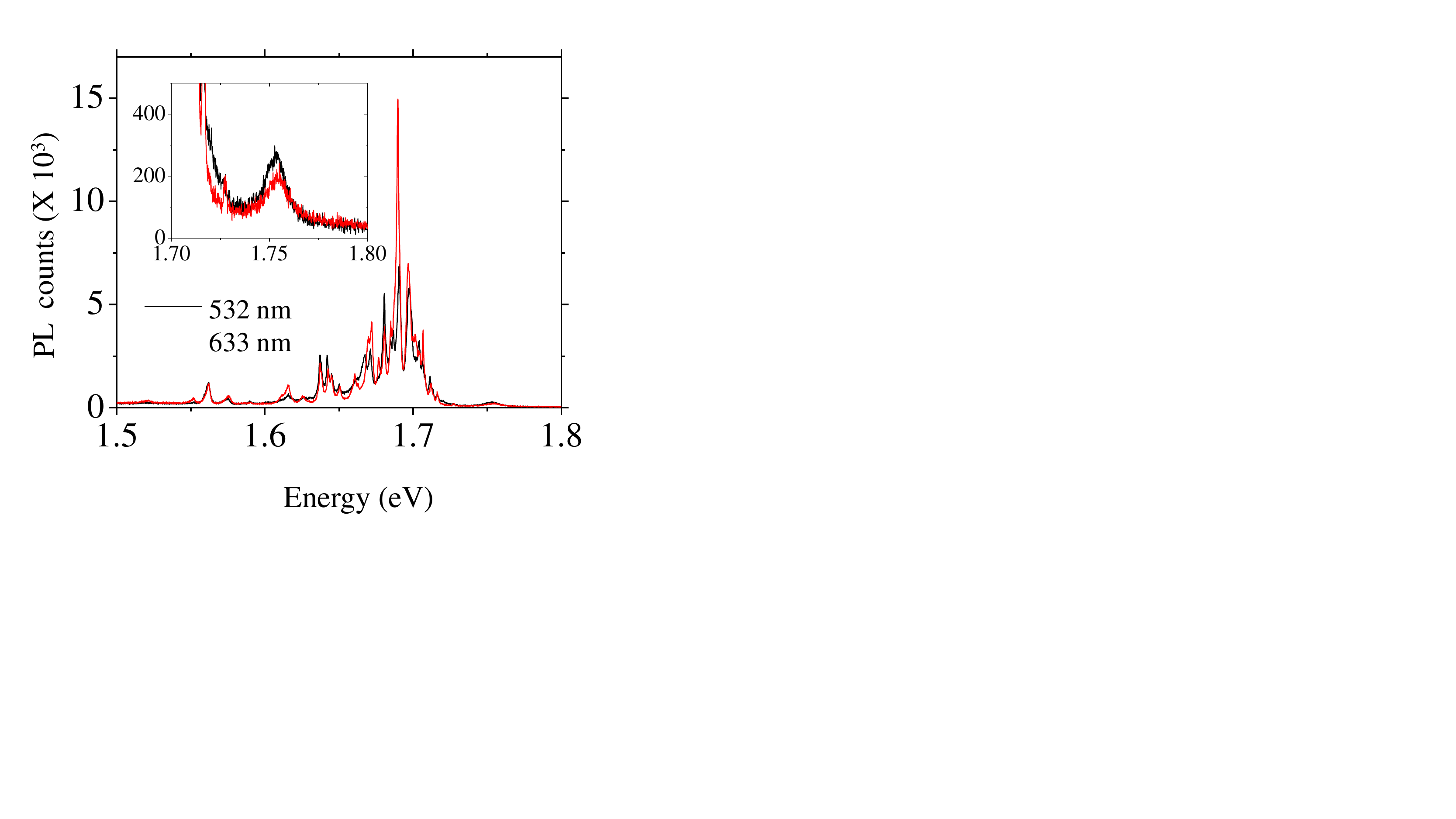}
	 	 \vspace{-1.5in}
	 	\caption{\textbf{Comparison of the PL spectra for 1L-WSe$_2$ on Au at 532 nm, 16$\mu$W excitation (black trace) and 633 nm, 5.8 $\mu$W excitation (red trace). Both spectra were acquired over an integration time of 30 seconds.}} \label{fig:S5}
	 \end{figure*}

	 \pagebreak

	 \section{Power dependent broadening of the defect features}
	 Figure S6a depicts PL spectra at different excitation power levels acquired from a particular spot of the 1L-WSe$_2$ flake on the same Au film as in figure 1a of the main text. Several sharp defect peaks can be observed even at higher powers. The PL spectra in the energy range of 1.65 eV to 1.72 eV have been illustrated in figure S6b, which clearly shows the sharp defect features in this range. Three such features namely D1, D2 and D3 have been highlighted in the black, red and brown dashed boxes respectively. A broadening of these defect features at higher laser excitation powers is conspicuous.

	 	 \begin{figure*}[!hbt]
	 	\includegraphics[scale=0.5] {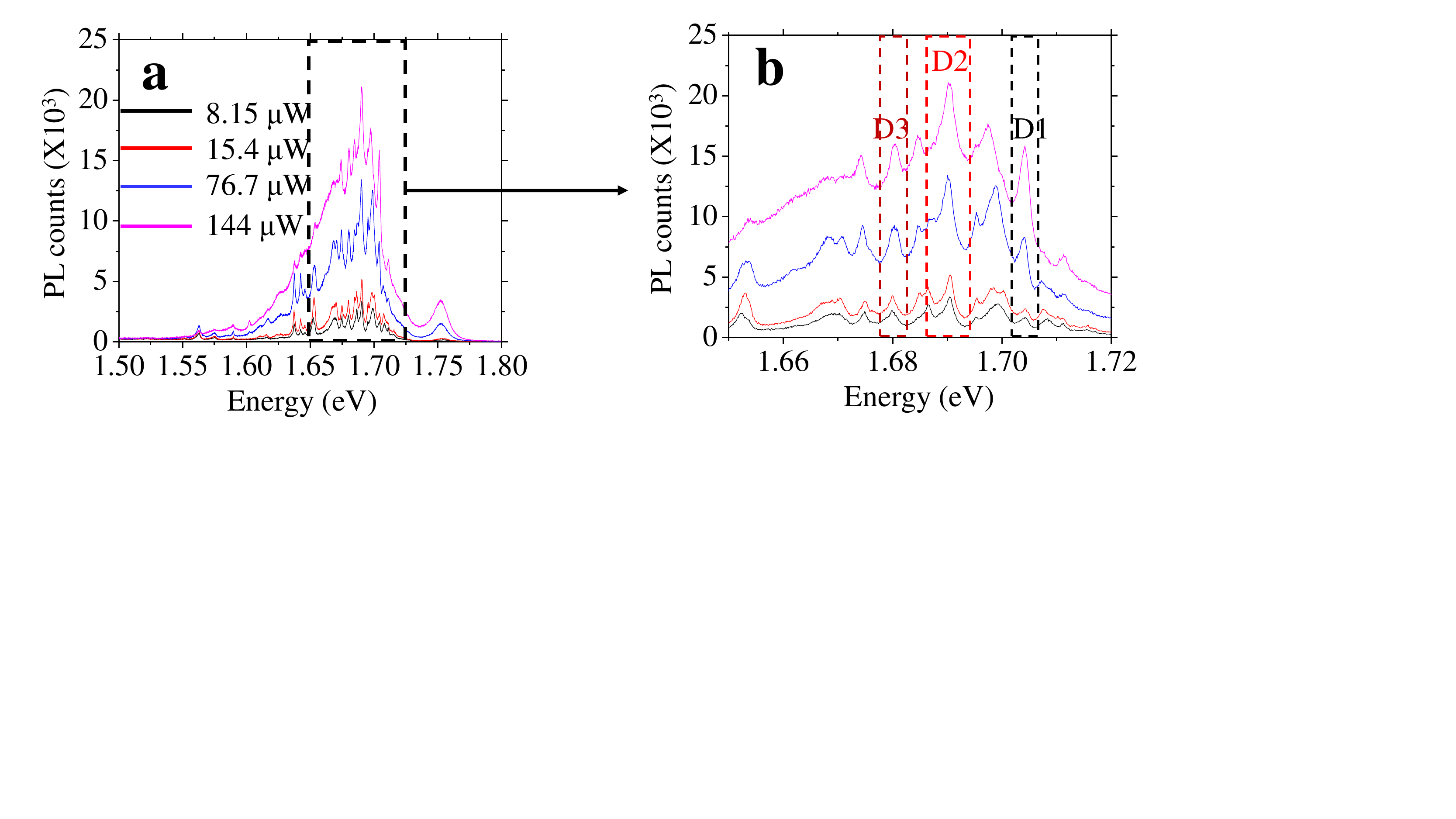}
	 	\vspace{-1.5in}
	 	\caption{(a) PL spectra accumulated at different excitation powers at 5 Kelvin, 532 nm excitation power, 30 seconds accumulation time. PL was acquired from a spot of the monolayer flake on the same Au film as in figure 1a of the main text. The dashed rectangle highlights the spectral region between 1.65 and 1.72 eV which has been depicted in figure S6b. (b) The spectra of figure S6a depicted in the spectral range of 1.65 eV to 1.72 eV. The sharp defect features are conspicuous at higher powers, and have been highlighted as D1, D2 and D3.} \label{fig:S6}
	 \end{figure*}

	 \pagebreak

	 \section{Comparison of PL spectra of 1L-WSe$_2$/hBN/SiO$_2$/Si and 1L-WSe$_2$/hBN/50 nm Au/SiO$_2$/Si}
	 Figure S7 shows comparison of PL spectra of 1L-WSe$_2$ on hBN/285 nm SiO$_2$/Si (control) and on hBN/50 nm Au/285 nm SiO$_2$/Si (stack) acquired at 633 nm, 5.8 $\mu$W excitation power over a 30 seconds acquisition time window. On the stack, it is observed that the entire spectrum is enhanced due to the fact that hBN acts as a very efficient barrier to the loss of excitons into the Au film by charge transfer and energy transfer due to its high band gap. However, from the point of view of the quantum emission purity, this enhancement is detrimental as the sharp peaks are now obscured by the background.
	
	 \begin{figure*}[!hbt]
	 	\includegraphics[scale=0.5] {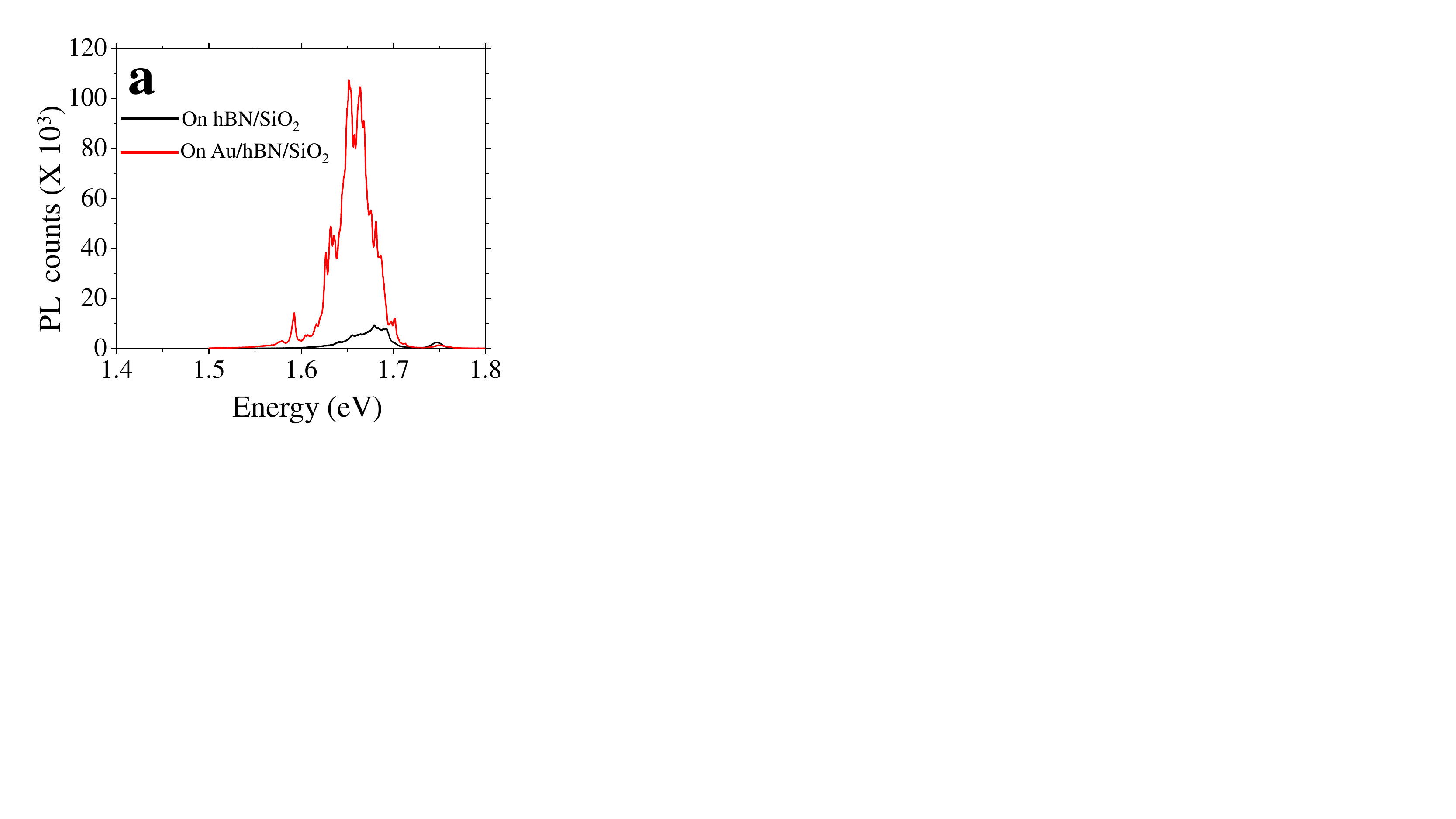}
	 	 \vspace{-2in}
	 	\caption{\textbf{PL spectra for 1L-WSe$_2$ on hBN/Au/SiO$_2$/Si and hBN/SiO$_2$/Si. Spectra acquired at 633 nm excitation, 5.8$\mu$W power and integration time of 30 seconds.} }\label{fig:S7}
	 \end{figure*}

	




\bibliographystyle{ieeetr}
\bibliography{Supporting_info}